\documentclass[a4paper,10pt,twoside,bibnote]{cpc-hepnp}
\usepackage{CJK,upgreek,fancyhdr}
\usepackage{multicol}
\usepackage{graphicx}
\usepackage{booktabs}
\usepackage{slashed}
\usepackage{amssymb,bm,mathrsfs,bbm,amscd}
\usepackage[tbtags]{amsmath}
\usepackage{lastpage}
\usepackage{multirow}
\usepackage{textcomp}
\usepackage{epstopdf}
\usepackage{float}
\usepackage{footnote}

\begin{document}
\begin{CJK*}{GBK}{song}

\setlength{\abovecaptionskip}{4pt plus1pt minus1pt}   
\setlength{\belowcaptionskip}{4pt plus1pt minus1pt}   
\setlength{\abovedisplayskip}{6pt plus1pt minus1pt}   
\setlength{\belowdisplayskip}{6pt plus1pt minus1pt}   
\addtolength{\thinmuskip}{-1mu}            
\addtolength{\medmuskip}{-2mu}             
\addtolength{\thickmuskip}{-2mu}           
\setlength{\belowrulesep}{0pt}          
\setlength{\aboverulesep}{0pt}          
\setlength{\arraycolsep}{2pt}           

\fancyhead[c]{\small Chinese Physics C~~~Vol. 42, No. 10 (2018) 103107}
\fancyfoot[C]{\small 103107-\thepage}
\footnotetext[0]{Received 31 May 2018, Published online 3 September 2018}

\title{Search for a lighter Higgs boson in the Next-to-Minimal Supersymmetric Standard Model\thanks{Supported by National Natural Science Foundation of China (11505208, 11661141007, 11705016, 11875275),
China Ministry of Science and Technology (2018YFA0403901) and partially by the France China Particle
Physics Laboratory (FCPPL) and CAS Center for Excellence in Particle Physics (CCEPP)}}

\author{
      Jun-Quan Tao$^{1;1)}$\email{taojq@mail.ihep.ac.cn}%
\quad M. Aamir Shahzad$^{1,2}$%
\quad Si-Jing Zhang$^{1,2,3}$%
\quad Chu Wang$^{1,2}$%
\\
      Yu-Qiao Shen$^{1,4}$%
\quad Guo-Ming Chen$^{1}$%
\quad He-Sheng Chen$^{1}$%
\quad S. Gascon-Shotkin$^{3}$%
\\
      M. Lethuillier$^{3}$%
\quad L. Finco$^{3}$%
\quad C. Camen$^{3}$%
}
\maketitle

\vspace{-1mm}
\address{
$^1$ Institute of High Energy Physics, Chinese Academy of Sciences, Beijing 100049, China \\
$^2$ University of Chinese Academy of Sciences, Beijing 100049, China \\
$^3$ Institut de Physique Nucl\'{e}aire de Lyon, Universit\'{e} de Lyon,\\ Universit\'{e} Claude Bernard Lyon 1, CNRS-IN2P3, Villeurbanne 69622, France \\
$^4$ Changzhou Institute of Technology, Changzhou 213032, China \\
}

\begin{abstract}
Following the discovery of the Higgs boson with a mass of approximately
125~GeV at the LHC, many studies have been performed from both the theoretical and experimental
viewpoints to search for a new Higgs Boson that is lighter than 125~GeV.
We explore the possibility of constraining a lighter neutral scalar Higgs boson $h_{1}$ and a lighter pseudo-scalar Higgs boson $a_{1}$ in the
Next-to-Minimal Supersymmetric Standard Model by restricting the
next-to-lightest scalar Higgs boson $h_{2}$ to be the one observed at the LHC after applying the phenomenological constraints and those from experimental measurements.
Such lighter particles are not yet completely excluded by the latest results of the search for a lighter Higgs boson in the diphoton decay channel from LHC data.
Our results show that some new constraints on the
Next-to-Minimal Supersymmetric Standard Model could be obtained for a lighter scalar Higgs boson at the LHC if such a search
is performed by experimental collaborations and more data.
The potentials of discovery for other interesting decay channels of such a lighter neutral scalar or pseudo-scalar particle are also discussed.
\end{abstract}


\begin{pacs}
11.30.Pb, 14.80.Cp     \qquad     {\bf DOI:} 10.1088/1674-1137/42/10/103107
\end{pacs}

\footnotetext[0]{\hspace*{-6mm}\raisebox{-1.05ex}{\includegraphics[scale=0.35]{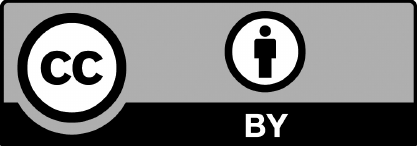}}
Content from this work may be used under the terms of the Creative
Commons Attribution 3.0 licence. Any further distribution of this
work must maintain attribution to the author(s) and the title of
the work, journal citation and DOI. Article funded by SCOAP$^3$
and published under licence by Chinese Physical Society and the
Institute of High Energy Physics of the Chinese Academy of
Sciences and the Institute of Modern Physics of the Chinese Academy of Sciences and IOP Publishing Ltd}%

\vspace{-1mm}
\begin{multicols}{2}

\section{Introduction}

The Standard Model (SM) of particle physics has been highly successful
in explaining high-energy experimental results. A Higgs boson with a mass of approximately
125~GeV was observed at the LHC with properties consistent with the
Higgs boson predicted by the SM~\cite{Aad:2012tfa,Chatrchyan:2012xdj,Aad:2013wqa,Chatrchyan:2013lba}.
However, the observed signal strength of the Higgs boson is somewhat biased against the
SM prediction within 1 or 2 times of the experimental uncertainty in each final state.
Many important questions dealing with the nature and origin of the Higgs boson
observed at the LHC remain unanswered.
The Higgs boson can be embedded in some models beyond the Standard Model (BSM), such as
supersymmetric models, which can easily accommodate
the discovered Higgs boson and address the deficiencies of the Standard Model.

Supersymmetry (SUSY)~\cite{Fayet:1976et,Fayet:1977yc,Farrar:1978xj,Giudice:1998xp} is one
theoretical possibility for BSM physics. It introduces a 
symmetry between fermions and bosons. The most common framework and minimal realization of SUSY
is the Minimal Supersymmetric Standard Model
(MSSM)~\cite{Nilles:1983ge,Haber:1984rc,Barbieri:1987xf}, which keeps limit the
number of new fields and couplings to the minimum. In the MSSM,
the Higgs sector contains two Higgs doublets, which leads to a spectrum
including two CP-even, one CP-odd, and two charged Higgs bosons. The
Lagrangian of the MSSM contains a supersymmetric mass term, namely the
$\mu$-term. This mass term is invariant under supersymmetry, and
therefore it seems unrelated to the electroweak scale, although it is\vspace{-4.3mm}\linebreak

\end{multicols}
\begin{multicols}{2}
\noindent
phenomenologically required to be in this scale. This leads to
the well-known ``$\mu$ problem"~\cite{Kim:1983dt,Giudice:1988yz} in
the MSSM. In addition, the MSSM suffers from another serious flaw, namely
the little hierarchy problem~\cite{Weinberg:1978ym,LlewellynSmith:1981yi}.
The simplest solution is the so-called
Next-to-Minimal Supersymmetric Standard Model (NMSSM),
which introduces a new gauge singlet superfield that only couples to
the Higgs sector in a similar manner to the Yukawa coupling, and can
generate a $\mu$ parameter dynamically of the order
of the SUSY breaking scale, solving the ``$\mu$ problem" and the little hierarchy problem without requiring much fine-tuning~\cite{Ellwanger:2009dp,Ellwanger:2011aa,Gunion:2012gc,King:2012is,Cao:2012fz,Agashe:2012zq,Kowalska:2012gs,Gherghetta:2012gb}.
Meanwhile, this new singlet adds additional degrees of freedom to
the NMSSM particle spectrum. In the CP conserving case, which is
assumed in this paper, the seven observable states in the Higgs sector can be
classified into three CP-even Higgs bosons $h_{i}$ ($i$ = 1,2,3), two
CP-odd Higgs bosons $a_{j}$ ($j$ = 1,2), and two charged Higgs bosons
$h^{\pm}$.

The extended parameter space of the NMSSM gives rise to a rich and
interesting phenomenology. The lightest CP-even Higgs bosons ($h_{1}$) with a mass range down to
approximately 80~GeV, assuming the next-to-lightest
CP-even Higgs boson $h_{2}$ as the new particle observed with a mass of
$\sim$125~GeV, was studied with the diphoton ($\gamma\gamma$) final state in~\cite{Jia-Wei:2013eea}.
A Higgs decaying into $\gamma\gamma$ is one of the two most promising channels for Higgs discovery at the LHC.
The discovery prospects for a light scalar in the NMSSM~\cite{Ellwanger:2015uaz}
and Two-Higgs Doublet Models (2HDM)~\cite{Cacciapaglia:2016tlr} have been considered, and comparisons with the CMS low mass diphoton analysis results at
$\sqrt{s}$ = 8 TeV~\cite{CMS:2015ocq} have been performed. Recently, the CMS collaboration
updated the results of the search for low mass Higgs bosons in the diphoton channel with the full 2016 data set
at $\sqrt{s}$ = 13 TeV~\cite{CMS:2017yta}. No significant excess has been observed by the CMS collaboration
in the mass range of 70~GeV to 110~GeV. The observed upper bounds on the corresponding signal rate may help to
place new constraints on the Next-to-Minimal Supersymmetric Standard Model.

In this paper, we explore the possibility of constraining a lighter neutral scalar Higgs boson $h_{1}$ and a lighter pseudo-scalar Higgs boson $a_{1}$ in the
NMSSM by restricting the next-to-lightest scalar Higgs boson $h_{2}$ to be the observed
125~GeV state, by comparing the lighter particles in the NMSSM with the latest CMS results with the full 2016 data set
at $\sqrt{s}$ = 13 TeV, after the constraints from the experimental measurements and other sources have been imposed.
The structure of this paper is as follows. In Section 2, 
we briefly introduce the Higgs sector of the NMSSM,
the details of different constraints, and
the chosen parameter ranges.
Section 3 
presents the results of the study for the lighter scalar Higgs boson $a_{1}$.
Section 4 
is dedicated to the study of the case in which the lighter resonance is the pseudo-scalar particle $a_{1}$.
Finally, the conclusions are presented in Section 5. 

\section{NMSSM and constraints on the NMSSM} \label{sec:nmssm}

\subsection{Description of NMSSM}
The general NMSSM is a supersymmetric extension of the Standard Model that
includes two Higgs superfields $\hat{H_{u}}$ and $\hat{H_{d}}$ and an additional gauge singlet chiral superfield
$\hat{S}$. 
In this paper, we consider the NMSSM with a scale invariant
superpotential $W_{\rm NMSSM}$ and the corresponding soft SUSY-breaking
masses and couplings $L_{soft}$, both of which are limited to the
R-parity and CP-conserving case. The superpotential $W_{\rm NMSSM}$,
depending on the Higgs superfields $\hat{H_{u}}$ and $\hat{H_{d}}$ and
$\hat{S}$, can be expressed as~\cite{Ellwanger:2009dp}
\begin{align}
\label{eq1}
W_{\rm NMSSM} =& h_{u}\hat{Q}\cdot\hat{H_{u}}\hat{U^{c}_{R}} + h_{d}\hat{H_{d}}\cdot\hat{Q}\hat{D^{c}_{R}} + h_{e}\hat{H_{d}}\cdot\hat{L}\hat{E^{c}_{R}} \nonumber\\
            & + \lambda\hat{S}\hat{H_{u}}\cdot\hat{H_{d}} +
            \frac{1}{3}\kappa\hat{S^{3}}.
\end{align}

\begin{figure*}[!b]
\centering
\includegraphics[width=0.3\textwidth,height=0.3\textwidth]{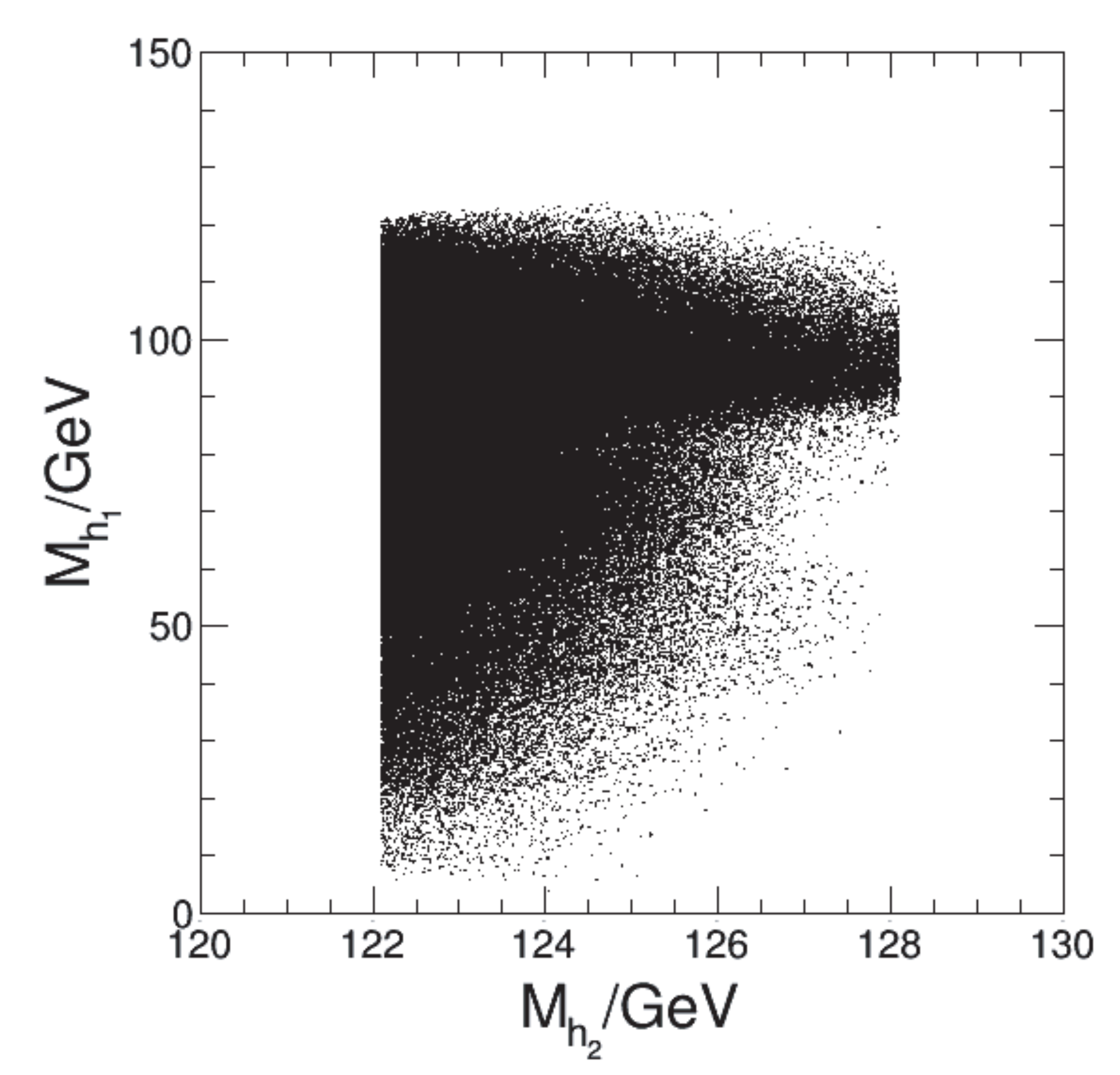}~~~
\includegraphics[width=0.3\textwidth,height=0.3\textwidth]{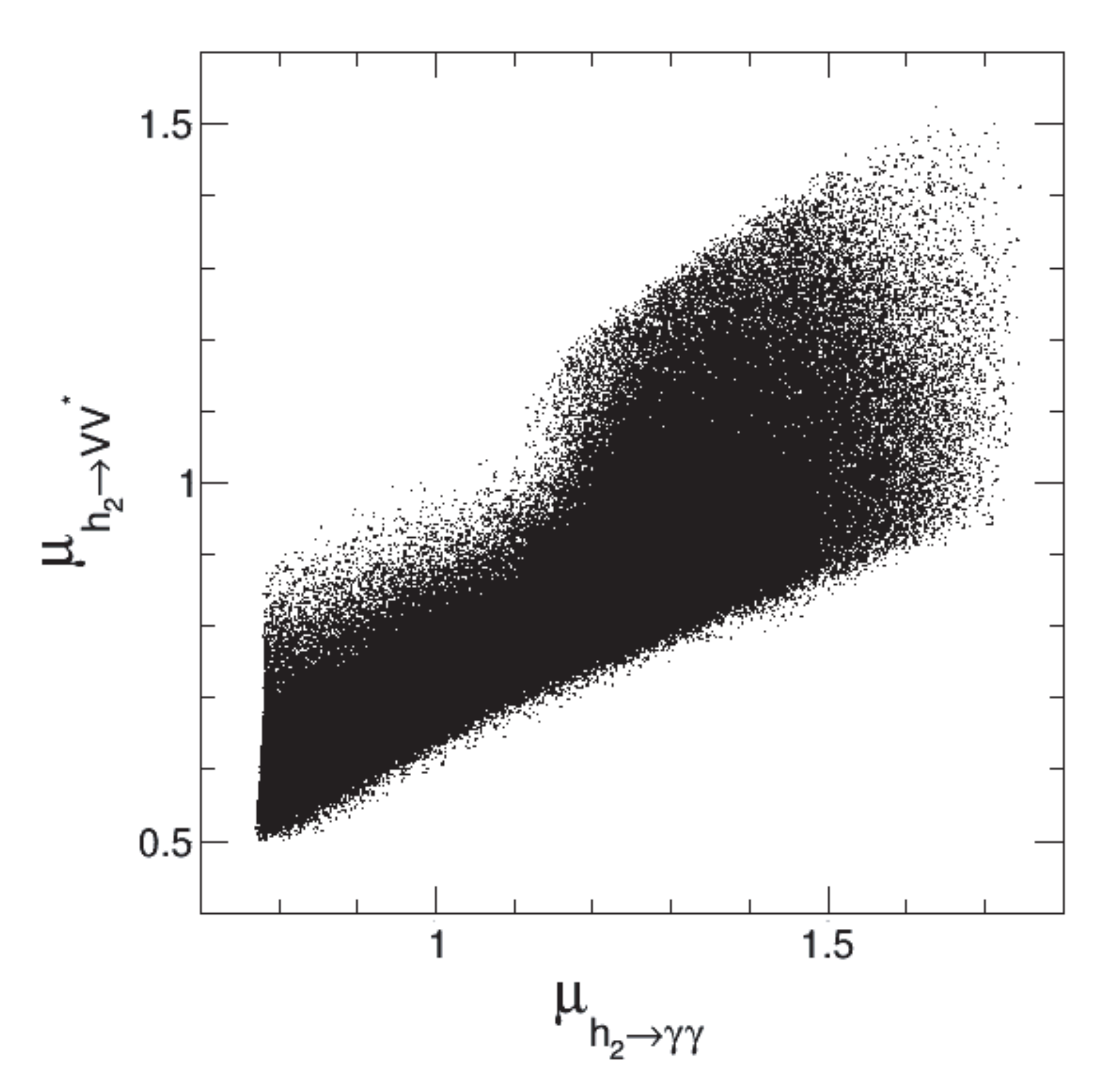}
\figcaption{\label{fig:h2mass_mu} Mass
spectrum of the two lightest scalar Higgs bosons $h_{1}$ and $h_{2}$ (left panel), and the signal strengths of $h_{2}\rightarrow\gamma\gamma$ versus the signal
strengths of $h_{2}\rightarrow VV^{*}$ (right panel), from the NMSSM scans with the constraints. }
\end{figure*}

In above formula, the first three terms on the right-hand side
represent the Yukawa couplings of the quark and lepton superfields. The
fourth term indicates that the $\mu$-term $\mu\hat{H_{u}}\hat{H_{d}}$ of
the MSSM superpotential is replaced by $\lambda\hat{S}\hat{H_{u}}\hat{H_{d}}$.
The last term, which is cubic in the singlet
superfield $\hat{S}$, is introduced to avoid the appearance of a Peccei-Quinn
axion, which is tightly constrained by cosmological
observations~\cite{Ellwanger:2009dp}. The corresponding soft SUSY-breaking
masses and couplings are given in the SLHA2~\cite{Allanach:2008qq}
conventions by the following equation~\cite{Ellwanger:2009dp}:
\begin{align}
\label{eq2}
-L_{\rm soft} =& m_{H_{u}}^{2}|H_{u}|^{2} + m_{H_{d}}^{2}|H_{d}|^{2} + m_{S}^{2}|S|^{2} + m_{Q}^{2}|Q|^{2}\nonumber\\
           &+ m_{U}^{2}|U_{R}|^{2} + m_{D}^{2}|D_{R}|^{2} + m_{L}^{2}|L|^{2} + m_{E}^{2}|E_{R}|^{2} \nonumber\\
           &+ h_{u}A_{u}Q \cdot H_{u}U_{R}^{c} - h_{d}A_{d}Q \cdot H_{d}D_{R}^{c} - h_{e}A_{e}L \cdot H_{d}E_{R}^{c} \nonumber\\
           &+ \lambda A_{\lambda}H_{u}\cdot H_{d}S + \frac{1}{3}\kappa A_{\kappa}S^{3} + {\rm h.c.}.
\end{align}

In the above Eq.~\ref{eq1} and Eq.~\ref{eq2}, it is clear that the non-zero vacuum
expectation value $s$ of the singlet $\hat{S}$ of the order of the
weak or SUSY-breaking scale gives rise to an effective $\mu$-term
with
\begin{equation}
\label{sixpar}
\mu_{\rm eff} = \lambda s .
\end{equation}
Here, $\lambda$ is dimensionless, hence the ``$\mu$ problem" of the MSSM is solved.
Meanwhile, the three SUSY-breaking mass-squared terms for $H_{u}$, $H_{d}$, and $S$
in $L_{\rm soft}$ can be expressed as functions of 
their VEVs (vacuum expectation values) through the three minimization conditions
of the scalar potential. Therefore, the Higgs sector of the NMSSM is
described by the following six parameters:
\begin{equation}
\lambda, \kappa, A_{\lambda}, A_{\kappa}, \tan{\beta}=\frac{\langle
H_{u}\rangle}{\langle H_{d}\rangle}, \mu_{\rm eff} = \lambda \langle S
\rangle,
\end{equation}
in which each pair of brackets denotes the VEV of the respective
superfield inside them. 
Besides these six\vspace{-4.3mm}\linebreak

\end{multicols}
\begin{multicols}{2}
\noindent
parameters of the Higgs
sector,
the squark and slepton
soft SUSY-breaking masses and the trilinear
couplings, as well as the gaugino soft SUSY-breaking masses, also must be specified, as described in the following section, in order to
describe the model completely.

\subsection{Constraints on the NMSSM and its parameters}

The program package NMSSMTools (version
5.2.0)~\cite{labNMSSMTools} is employed in this study 
to calculate the SUSY particles, the spectrums of the NMSSM Higgs bosons, their decaying branching ratios (BR),
and the reduced couplings of the NMSSM Higgs bosons to other particles.
NMSSMTools contains four subpackages,  NMHDECAY, NMSDECAY, NMSPEC, and
NMGMSB. The FORTRAN code NMHDECAY provides the masses, decay widths,
branching ratios and reduced couplings to other particles, for Higgs bosons that will be used in this paper.
In this paper, we consider the four Higgs production modes of gluon-gluon fusion through a heavy quark loop (ggh),
the vector boson fusion process (vbf), the associate production of Higgs with a vector boson (vh),
and the associated production of a Higgs with a pair of top quarks (tth).
The cross sections for different production modes of each NMSSM Higgs boson are obtained from the linear interpolation of
the 5 GeV per step cross section values taken from the handbook of the
LHC Higgs Cross Section Working Group~\cite{deFlorian:2016spz} for a SM-like BSM Higgs boson, and
multiplied by the reduced couplings of each NMSSM Higgs boson to gluons $\kappa_{g}^2$,
gauge bosons $\kappa_{V}^2$, and fermions $\kappa_{f}^2$,
which are given by the output of NMSSMTools.
The NMSSMTools package applies all phenomenological constraints, including the absence of Landau singularities
below the GUT scale and the constraints
from flavor physics, dark matter relic density $\Omega h^{2}$,
anomalous magnetic moment of the muon ($g-2$), Higgs searches in various channels, and direct searches for SUSY particles at
LEP, Tevatron, and LHC, with the details described in~\cite{labNMSSMTools}.

\begin{figure*}[!b]
\centering
\includegraphics[width=0.3\textwidth,height=0.3\textwidth]{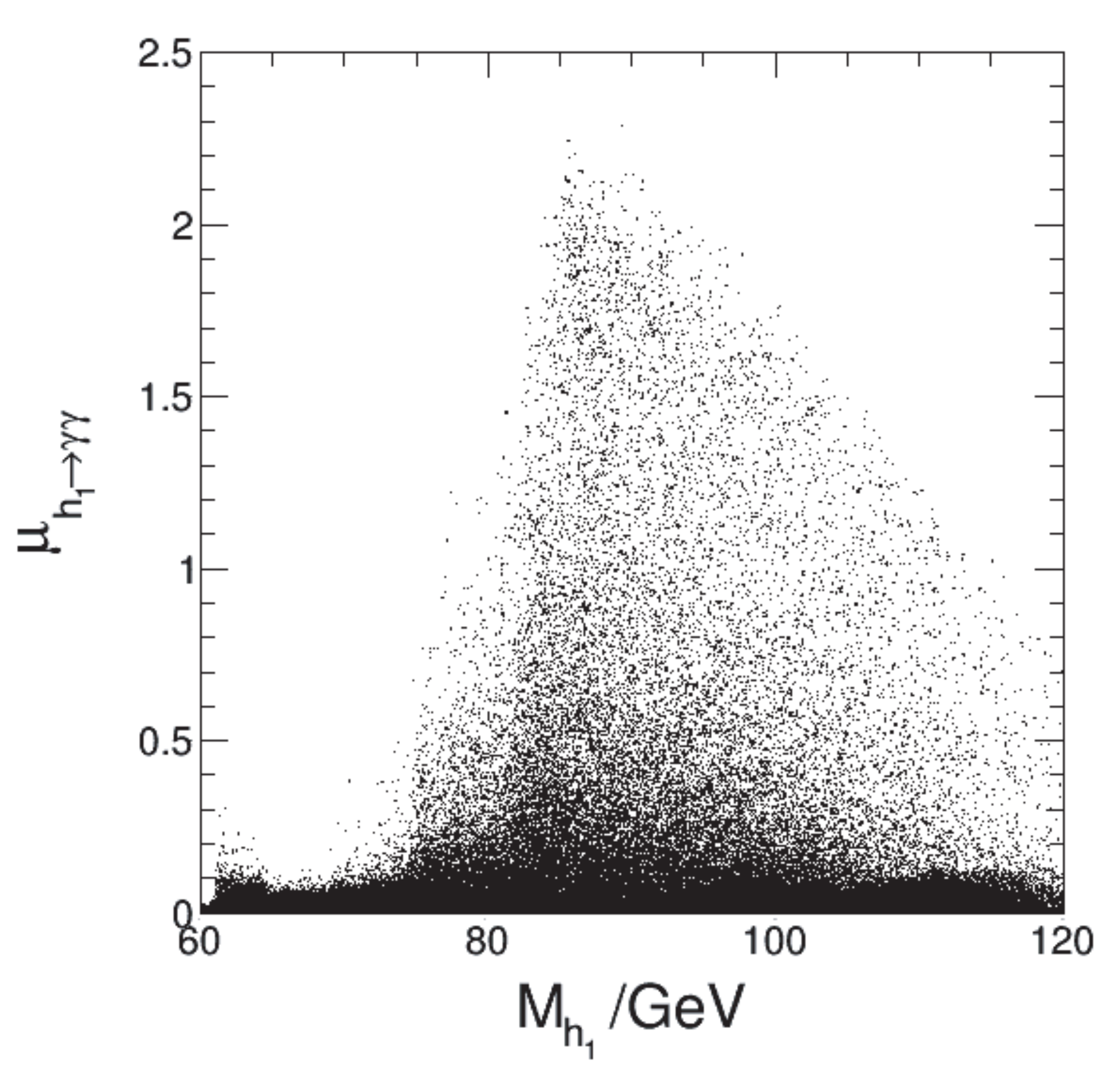}\quad
\includegraphics[width=0.3\textwidth,height=0.3\textwidth]{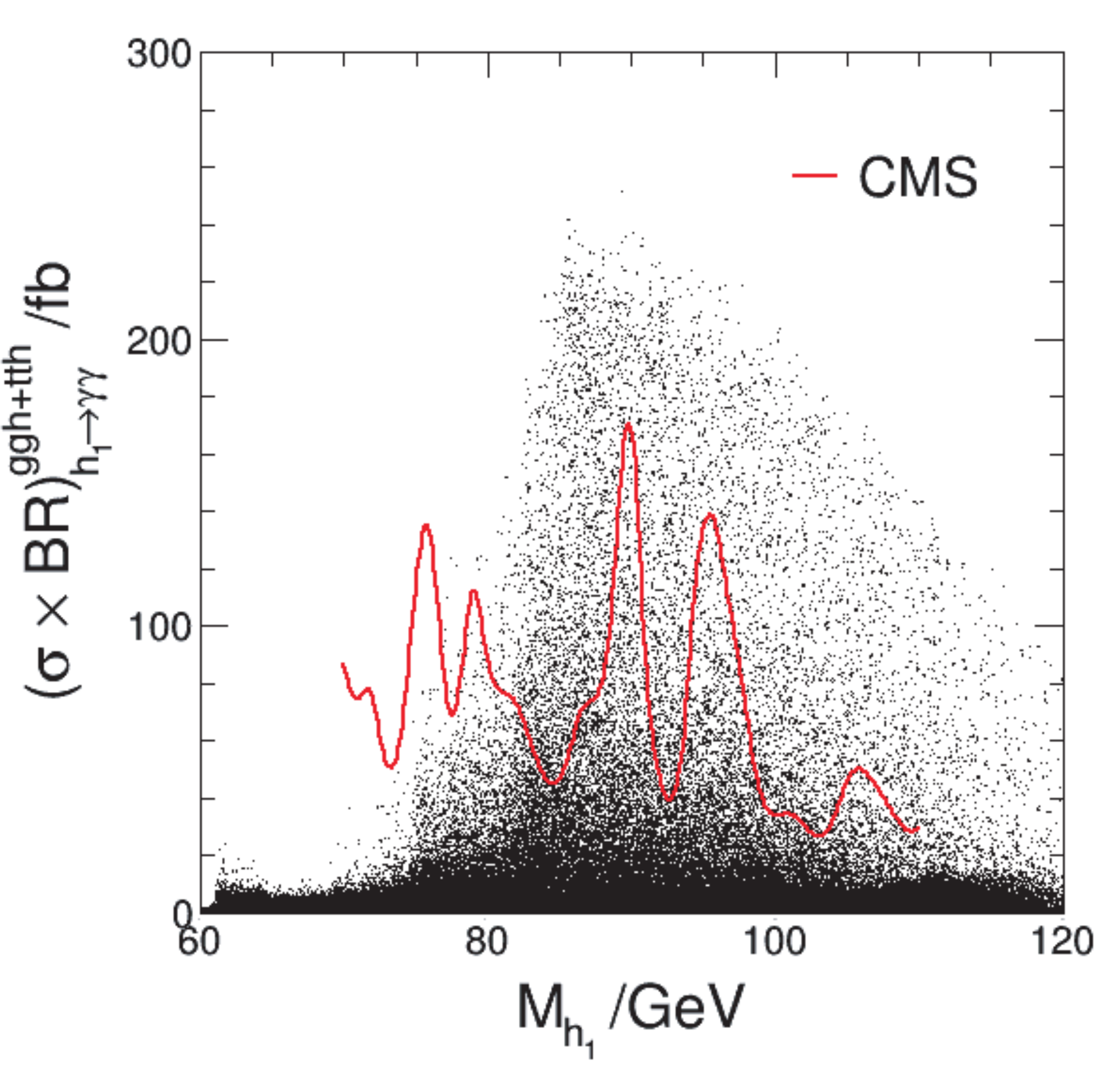}\quad
\includegraphics[width=0.3\textwidth,height=0.3\textwidth]{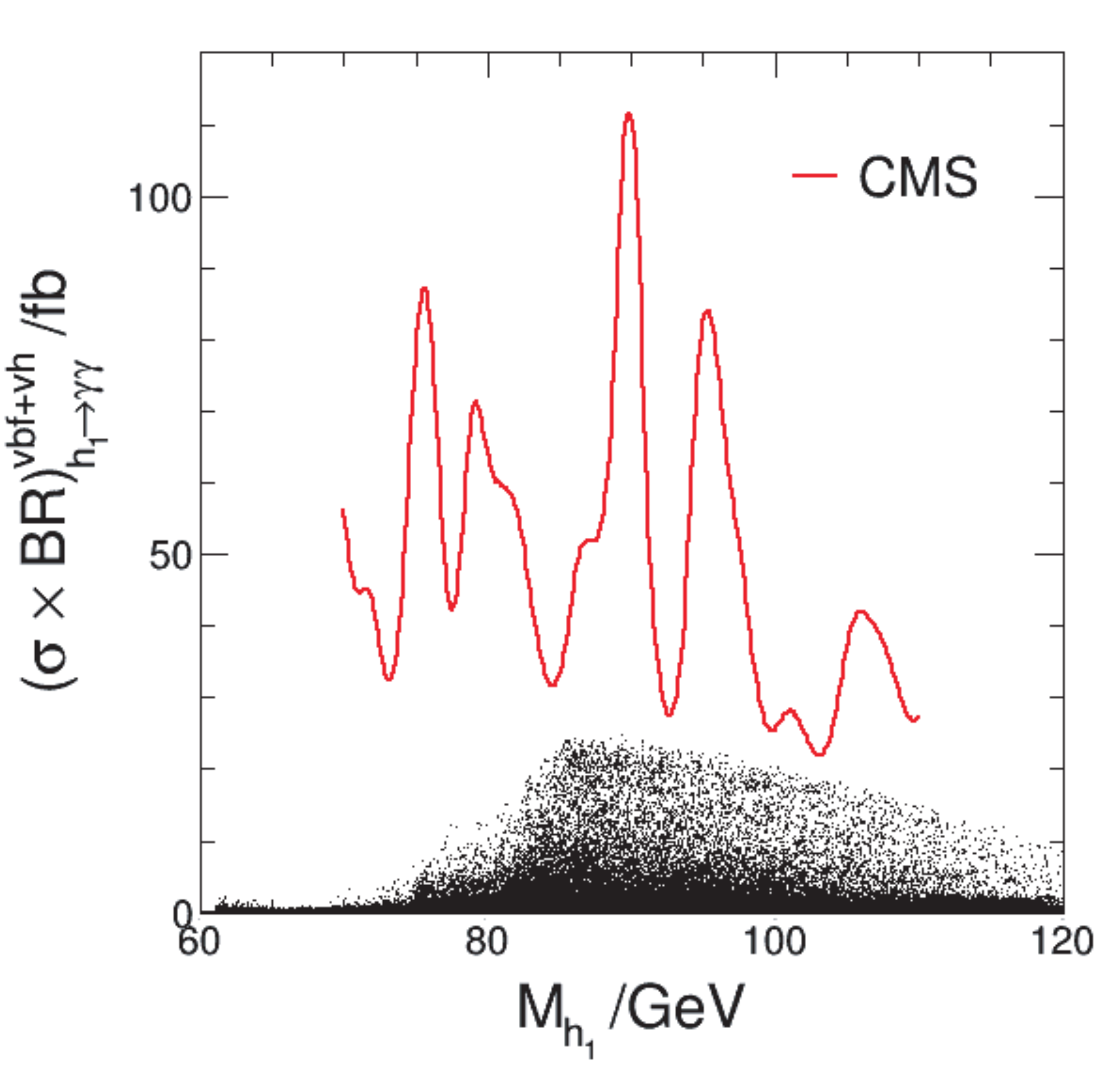}
\figcaption{\label{fig:h1ggmu_mass}(color online) Signal strengths of $h_{1}$ decaying into the diphoton ($\mu_{h_{1}\rightarrow\gamma\gamma}$) versus the mass of $h_{1}$ (left panel) and the signal rates as functions of the $h_{1}$ mass generated in the general NMSSM superimposed on the observed results of the CMS 13~TeV low-mass diphoton analysis~\cite{CMS:2017yta} in the combined ggh and tth production mode $(\sigma \times BR)_{h_{1}\rightarrow\gamma\gamma}^{ggh+tth}$ (middle panel) and in the combined vbf and vh production mode $(\sigma \times BR)_{h_{1}\rightarrow\gamma\gamma}^{vbf+vh}$ (right panel). }
\end{figure*}

The six NMSSM specific parameters described above are varied in the following ranges:
\begin{equation}
\label{sixparcons}
\begin{split}
0.55 < \lambda < 0.75, \quad 0.05 < \kappa < 0.3, \quad 3 < \tan\beta < 6 , \\
150\rm ~GeV < \mu_{eff} < 350\rm ~GeV, \\
-500\rm ~GeV < A_{\kappa} < 0\rm ~GeV, \\
500\rm ~GeV < A_{\lambda} < 1600\rm ~GeV.
\end{split}
\end{equation}

To choose values of $\lambda$ and $\kappa$ that are sufficiently large but
small enough to avoid a Landau pole below the GUT scale and low values for
$\tan\beta$ that naturally keep the amount of fine-tuning as low
as possible is suitable for this study.
We found that wider ranges of the trilinear couplings
$A_{\kappa}$, $A_{\lambda}$, and $\mu_{\rm eff}$ have practically no impact on our results.

The soft SUSY-breaking mass terms for the squark ($M_U$, $M_D$, and $M_Q$)
and sleptons ($M_L$ and $M_E$),
the soft SUSY-breaking trilinear couplings ($A_{D}$, $A_{E}$, and $A_{U}$),
the gluino mass ($M_3$), and the other soft SUSY-breaking gaugino masses ($M_1$ and $M_2$)
have been set as
\begin{equation}
\label{susyparcons}
\begin{split}
M_Q = M_U = M_D = 1000~{\rm GeV}, \\
M_L = M_E = 300~{\rm GeV}, \\
A_{D} = A_{E} = A_{U} = 1000~{\rm GeV}, \\
100~{\rm GeV} < M_{1} < 150~{\rm GeV}, \\
150~{\rm GeV} < M_{2} < 250~{\rm GeV}, \\
1000~{\rm GeV} < M_{3} < 2000~{\rm GeV}.
\end{split}
\end{equation}

Using the NMSSMTools package and the general NMSSM model, we have performed random scans with about 10 billion points in the specific parameter space described above.
For each point in the parameter space satisfying the phenomenological constraints,
we require that an SM-like Higgs state, the next-to-lightest scalar Higgs boson $h_{2}$ in NMSSM,
must be within the allowed theoretical uncertainty of 3~GeV around the measured mass 125.1 ~GeV
at the LHC using the whole Run1 data~\cite{Aad:2015zhl} ($125.1 \pm 3$~GeV)
and couplings of $h_{2}$  to gauge bosons and fermions in the $3\sigma$ ranges of the best-fit values
given in~\cite{Khachatryan:2016vau,Bernon:2014vta}. With these constraints, around 1.40 million points remain.
Fig.~\ref{fig:h2mass_mu} shows the mass
distributions of the two lightest scalar Higgs bosons $h_{1}$ and $h_{2}$, and the signal strengths of $h_{2}\rightarrow\gamma\gamma$ versus the signal
strengths of $h_{2}\rightarrow VV^{*}$ ($VV^{*}$=$ZZ$/$W^{+}W^{-}$), with the signal strength defined as the relative ratio of
the cross section time branching ratio of the Higgs boson from each NMSSM point to the SM-like BSM predicted
value from the handbook of the LHC Higgs Cross Section Working Group~\cite{deFlorian:2016spz}.

With the full 2016 data set at $\sqrt{s}$ = 13 TeV, the CMS collaboration have updated the results for searching for a light resonance decaying into two photons
in the mass range from 70~GeV to 110~GeV~\cite{CMS:2017yta}.
To compare with the experimental sensitivity
 and to explore the discovery potentials for other interesting decay channels,
the masses of the lightest scalar Higgs boson $h_{1}$ and the lightest pseudo-scalar Higgs boson $a_{1}$
are constrained in the range from 60~GeV to 120~GeV in the following sections.

\begin{figure*}[!t]
\centering
\includegraphics[width=0.3\textwidth,height=0.3\textwidth]{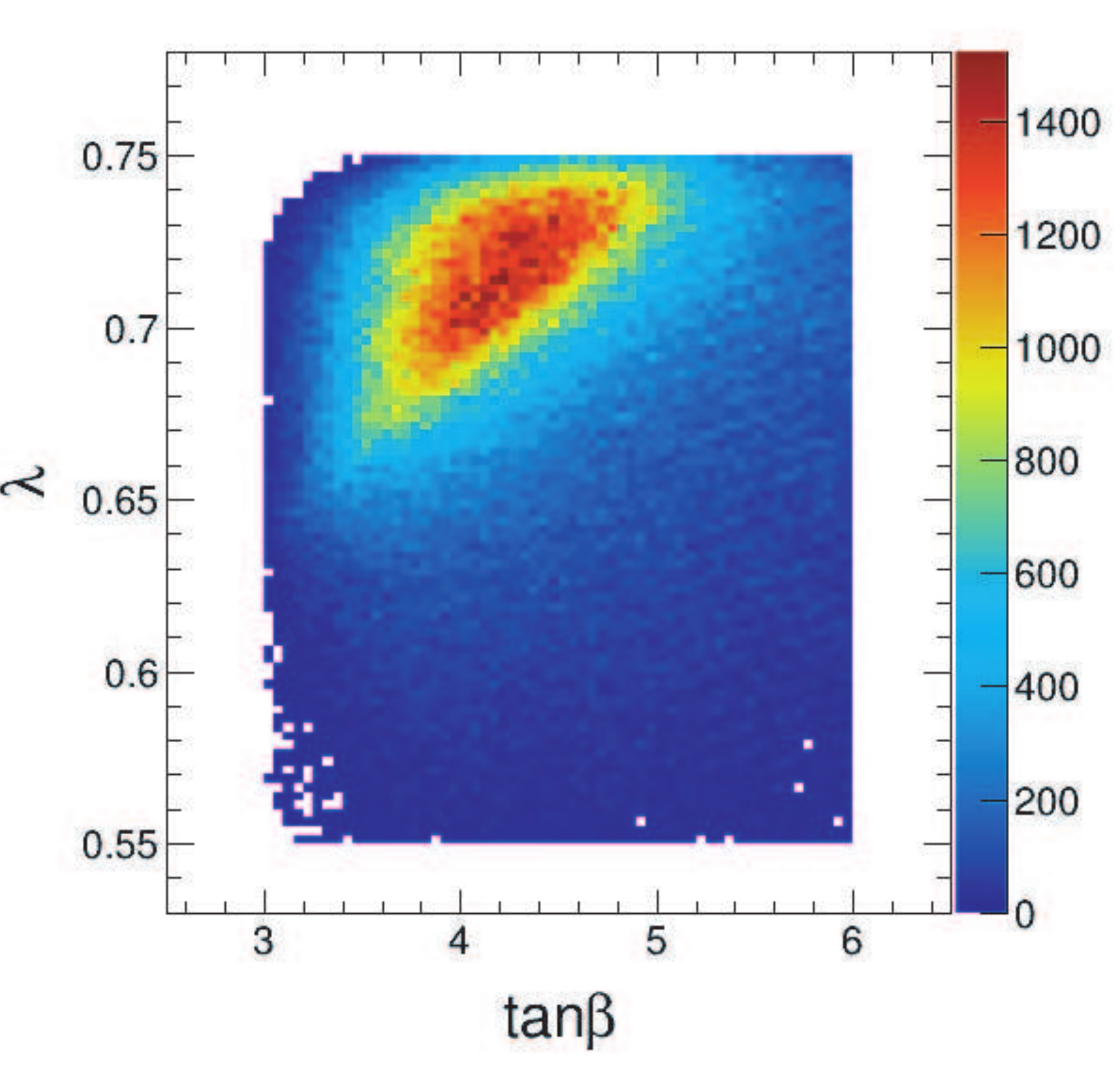}\quad
\includegraphics[width=0.3\textwidth,height=0.3\textwidth]{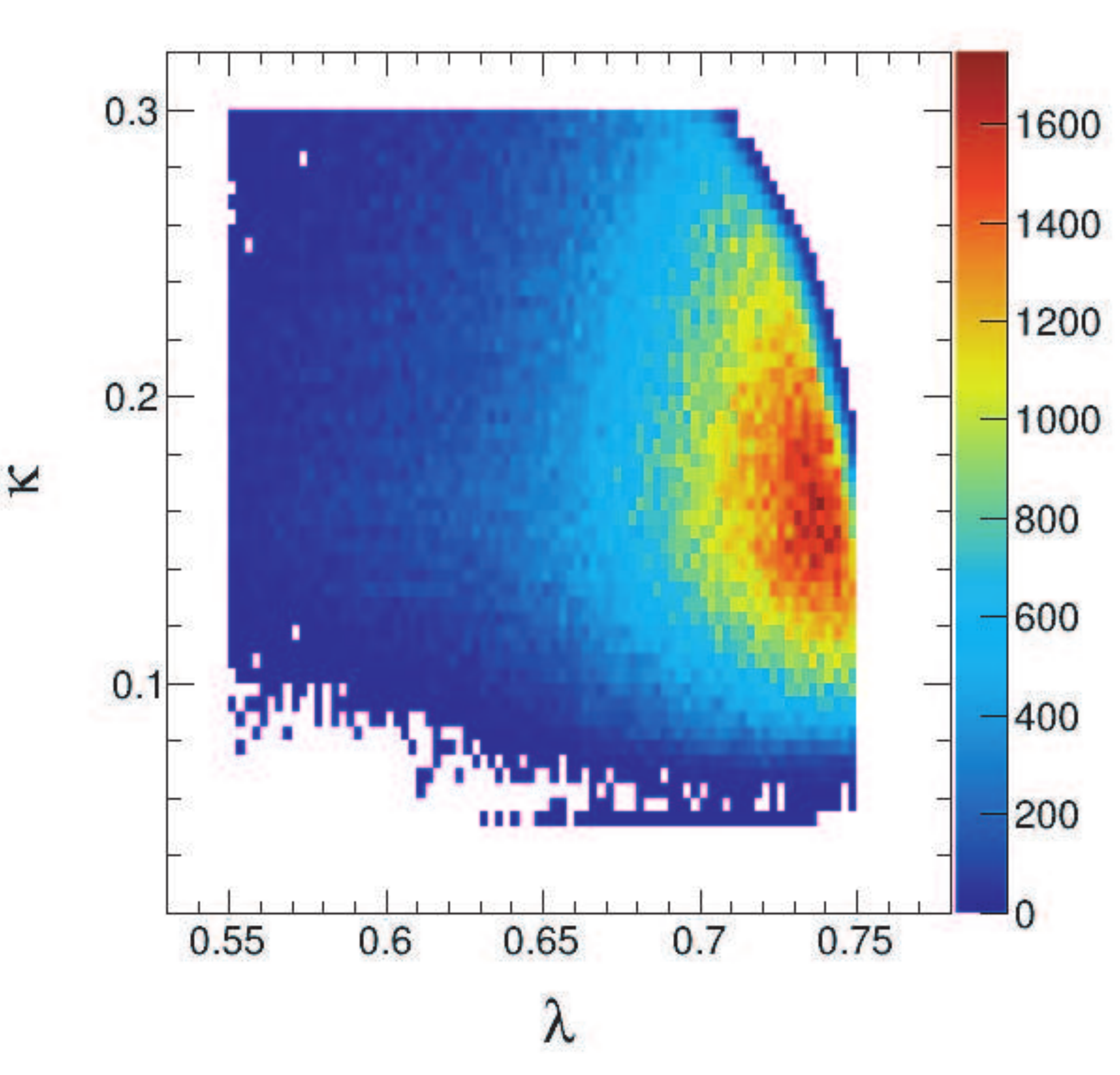}\quad
\includegraphics[width=0.3\textwidth,height=0.3\textwidth]{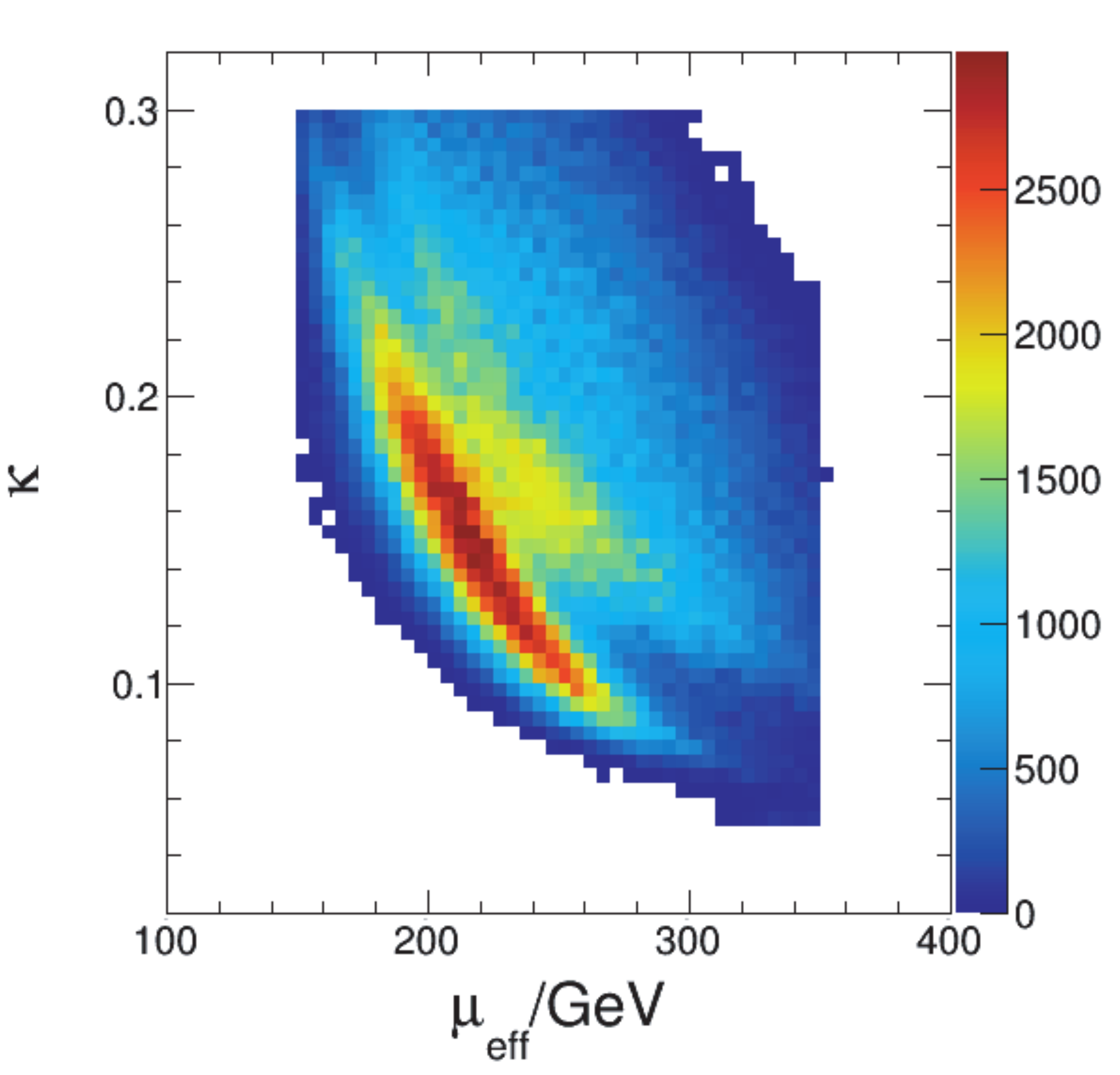}  \\
\includegraphics[width=0.3\textwidth,height=0.3\textwidth]{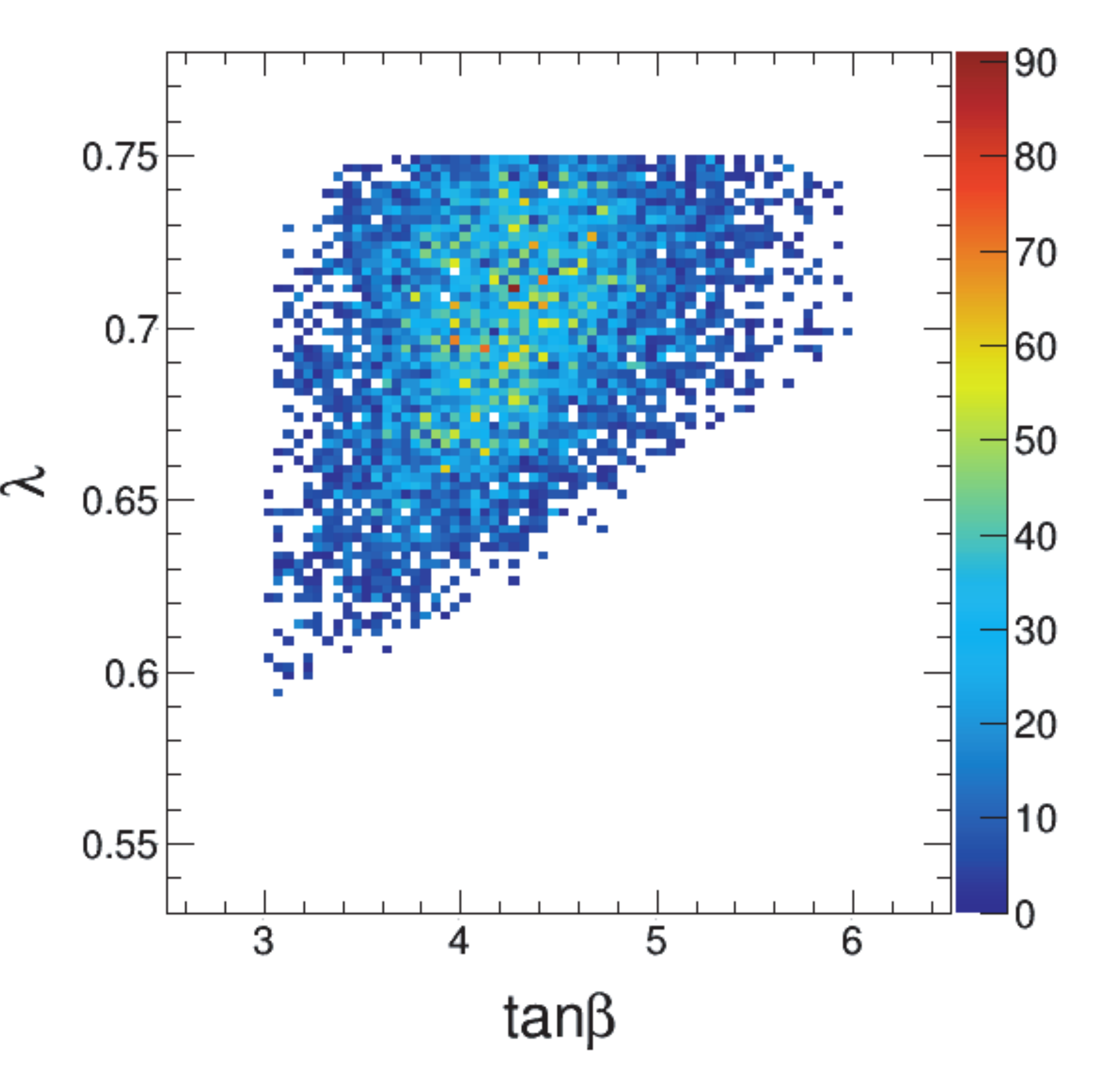}\quad
\includegraphics[width=0.3\textwidth,height=0.3\textwidth]{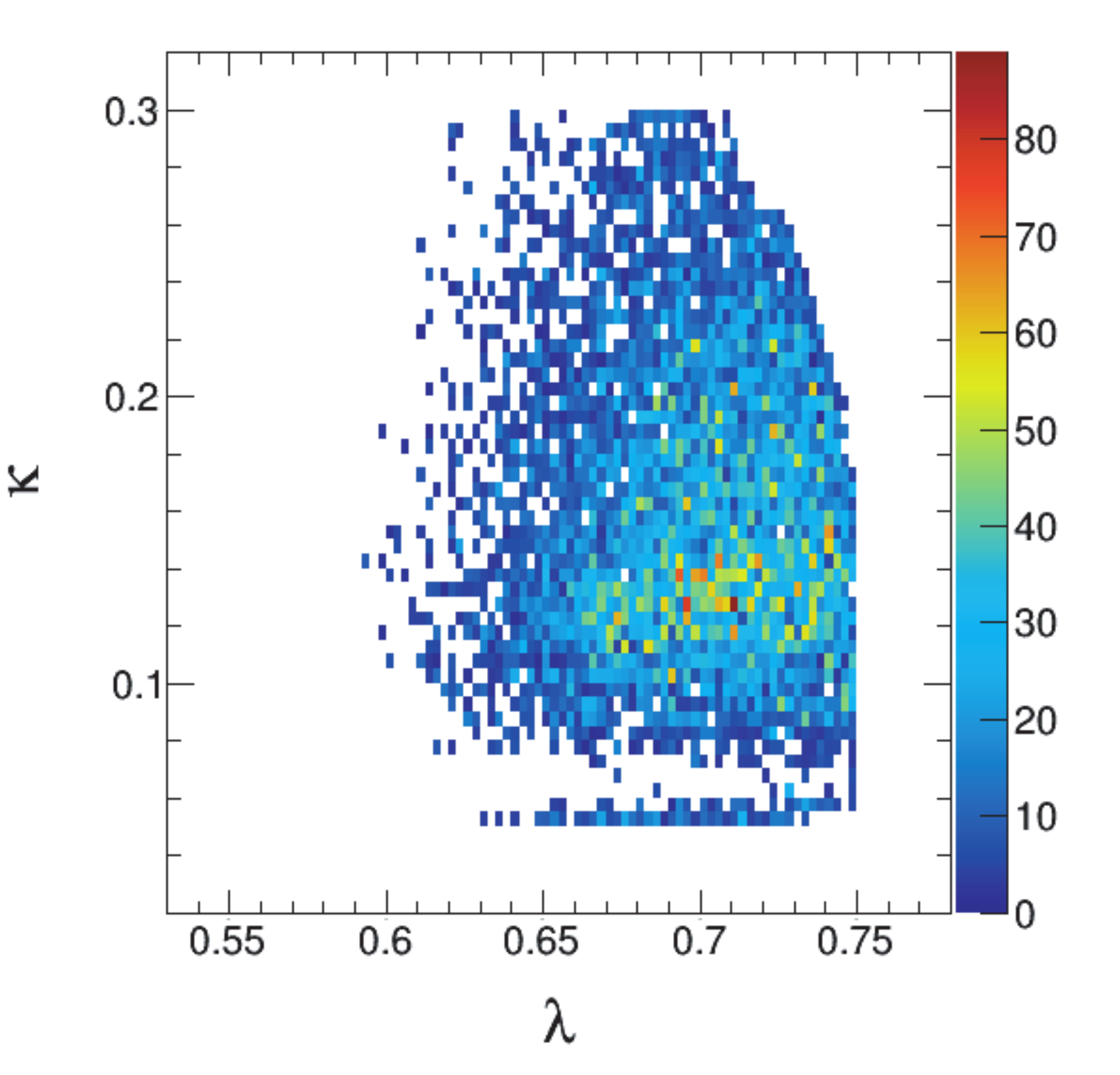}\quad
\includegraphics[width=0.3\textwidth,height=0.3\textwidth]{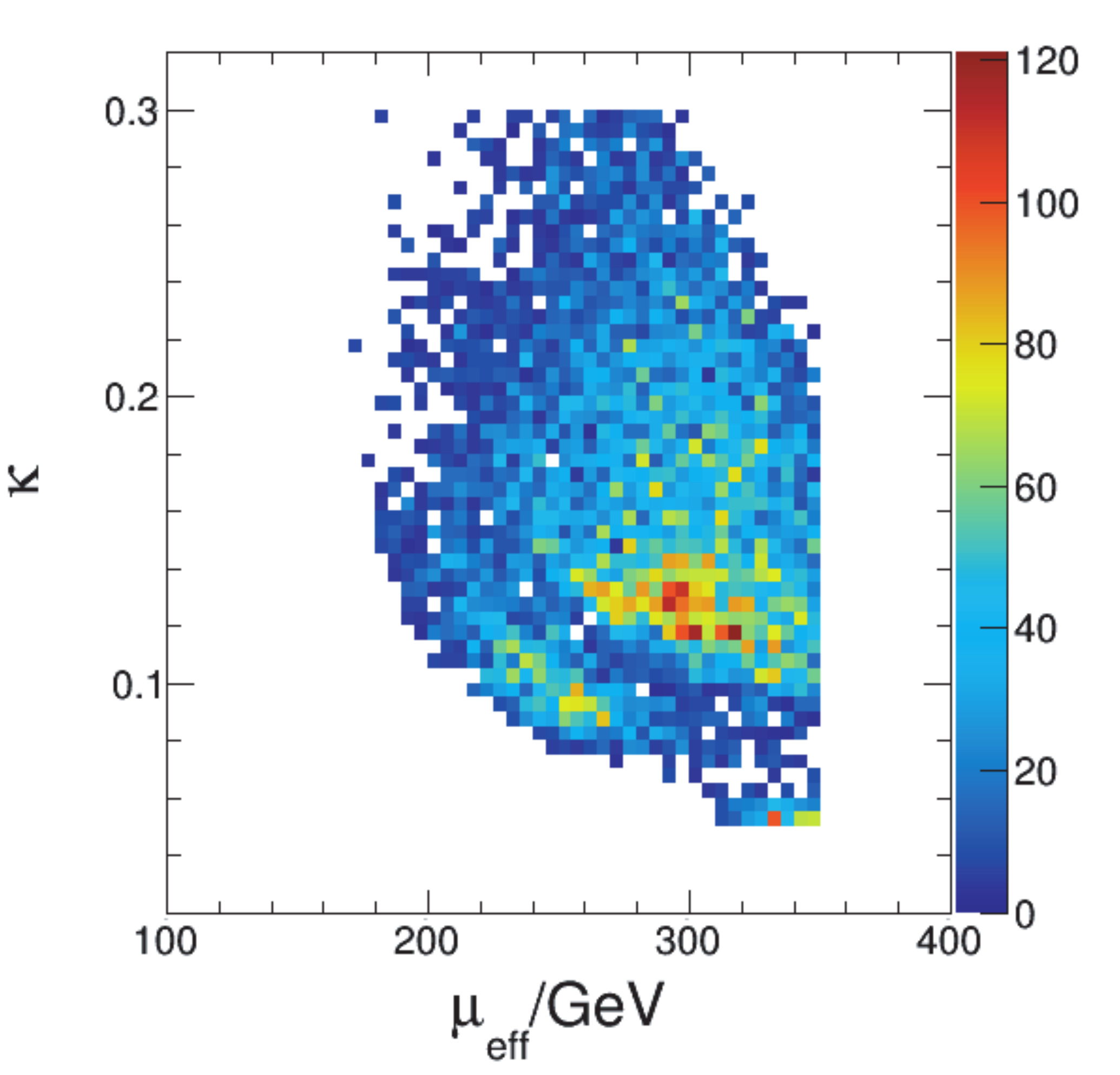}
\figcaption{\label{fig:h1IncExc2DPar} (color online) Two-dimensional scatter plots of the input parameters $\tan\beta$ vs $\lambda$, $\lambda$ vs $\kappa$, and $\mu_{\rm eff}$ vs $\kappa$ for all selected points passing the phenomenological constraints, the mass and signal strength constraints on $h_{2}$, and the mass constraint on $h_{1}$ with a mass range from 60~GeV to 120~GeV are shown in the top panels. The corresponding two-dimensional scatter plots of the input parameters for the selected points further excluded by the observed upper limits of the CMS 13~TeV low-mass diphoton analysis with a mass range from 70~GeV to 110~GeV are shown in the bottom panels. }
\vspace{1mm}
\end{figure*}

\section{Results for a lighter scalar Higgs boson} \label{sec:scalar}

In this section, we will explore the possibility that the signal may be given by the lightest scalar Higgs
boson $h_{1}$ in the NMSSM. We perform a detailed comparison with the sensitivity of the CMS search at 13 TeV.
About 1.25 million points are selected from the random scans passing the phenomenological constraints,
the mass and signal strength constraints on $h_{2}$, and the mass constraint on $h_{1}$ with a mass range from 60~GeV to 120~GeV .

The left panel of Fig.~\ref{fig:h1ggmu_mass} shows the signal strengths of $h_{1}$ decaying into a diphoton
$\mu_{h_{1}\rightarrow\gamma\gamma}$ plotted against the mass of the Higgs boson $h_{1}$
($M_{h_{1}}$). It can be seen that a sizable enhancement over the SM-like Higgs rate is
possible for the Higgs boson $h_{1}$, with the largest strength $\sim2.2$ occurring at an $h_{1}$ mass
of $\sim$85~GeV. We note that for the mass ranges $M_{h_{1}}$ $<$ $\sim$80~GeV and $M_{h_{1}}$ $>$ $\sim$110~GeV, the allowed signal strengths
$\mu_{h_{1}\rightarrow\gamma\gamma}$ are considerably lower than 1. In particular, for $M_{h_{1}}$ $<$ $\sim$75~GeV the signal strengths are below $\sim$0.2.
The production rates in femtobarns ($fb$) of $h_{1}$ decaying into $\gamma\gamma$ versus $M_{h_{1}}$ are also plotted for the combined ggh and
tth production mode ($(\sigma \times BR)_{h_{1}\rightarrow\gamma\gamma}^{ggh+tth}$) in the middle panel and for the combined vbf and vh production mode ($(\sigma \times BR)_{h_{1}\rightarrow\gamma\gamma}^{vbf+vh}$) in the right panel of Fig.~\ref{fig:h1ggmu_mass},
superimposed on the public observed exclusion limits of the CMS collaboration with the full 2016 data set at $\sqrt{s}$ = 13 TeV~\cite{CMS:2017yta} shown by the red line.
These comparisons show that there is no sensitivity in the vbf+vh production mode, but many points are above the CMS observed upper limit in the ggh+tth production mode for a light Higgs boson with mass $M_{h_{1}}$ $>$ $\sim$80~GeV. For the vbf+vh production mode, it is possible to obtain points above the CMS observed upper limit in the near future with more
proton-proton collision data accumulated at the LHC.

As the points above the observed CMS upper limit are excluded at the 95\% confidence level, we can expect to exclude some NMSSM region in the parameter space thanks to this
analysis. To illustrate this point, in Fig.~\ref{fig:h1IncExc2DPar} we compare several sensitivity parameters for two different cases.
The top three figures show all the selected 1.25 million points in the 2D planes of $\tan\beta$ vs $\lambda$, $\lambda$ vs $\kappa$, and $\mu_{\rm eff}$ vs $\kappa$.
The bottom panel shows the corresponding figures for the points with $(\sigma \times BR)_{h_{1}\rightarrow\gamma\gamma}$ above the CMS observed upper limit,
and consequently excluded by the experiment in the mass range from 70~GeV to 110~GeV. The parameter ranges of the excluded points overlap with the rest points that are not excluded by the experiment.
Therefore, we cannot conclude that the parameter ranges shown in the bottom three figures are excluded by the experiment.
Nevertheless, it can be seen by comparing of the top and bottom figures that the points with higher $\tan\beta$ but lower $\lambda$ from the bottom-left panel, with lower $\lambda$ ($<$ 0.6) from the bottom-middle panel, or with lower $\mu_{\rm eff}$ ($<$ 180~GeV) from the bottom-right panel
are not sufficiently sensitive to larger production rates to be excluded by the experiment.
The excluded points populate the parameter space with $\tan\beta$ around 4, $\lambda$ around 7,  $\kappa$ around 0.13, and $\mu_{\rm eff}$ around 300~GeV.

\begin{figure*}[!th]
\centering
\includegraphics[width=0.3\textwidth,height=0.3\textwidth]{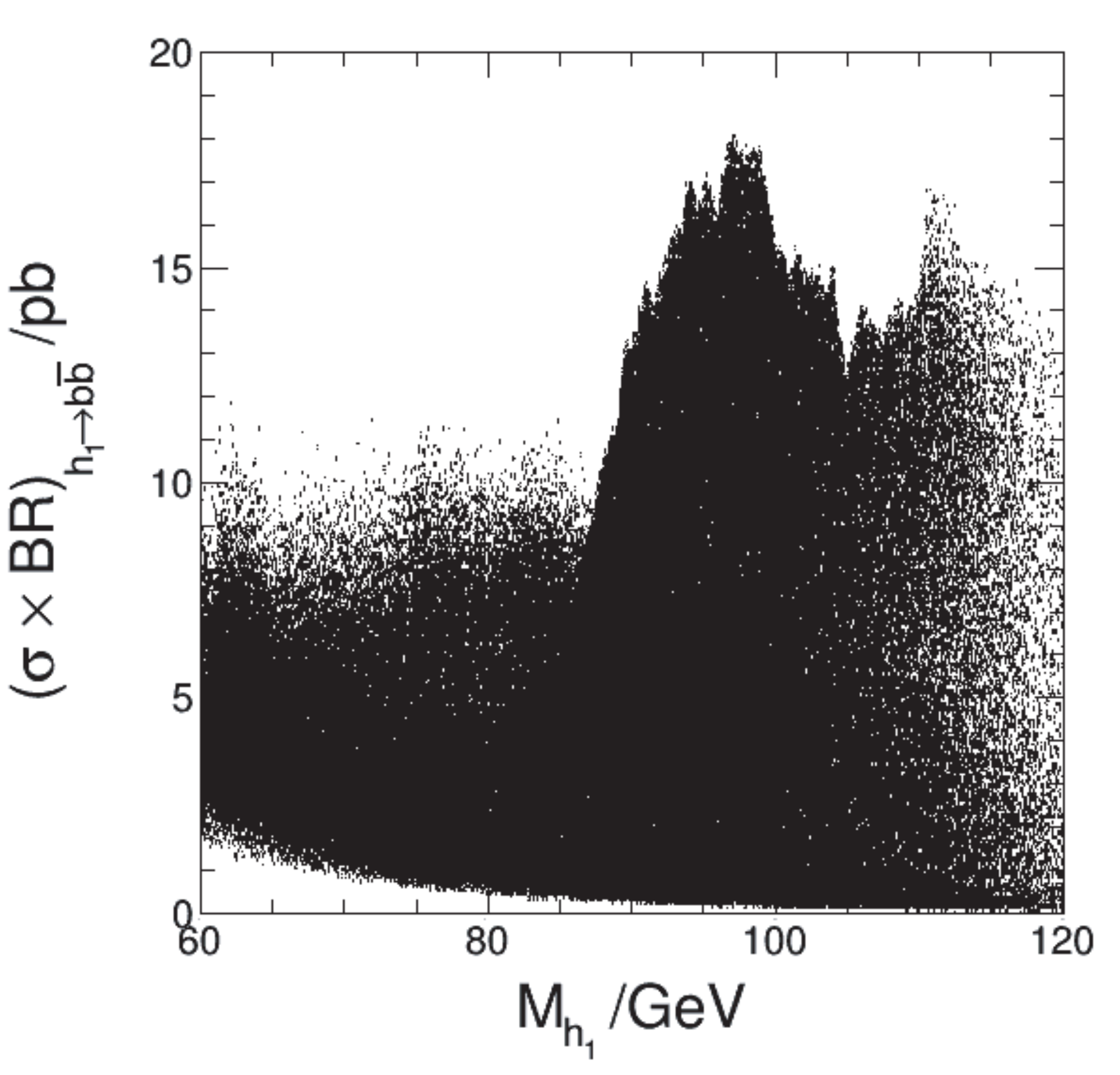}\quad
\includegraphics[width=0.3\textwidth,height=0.3\textwidth]{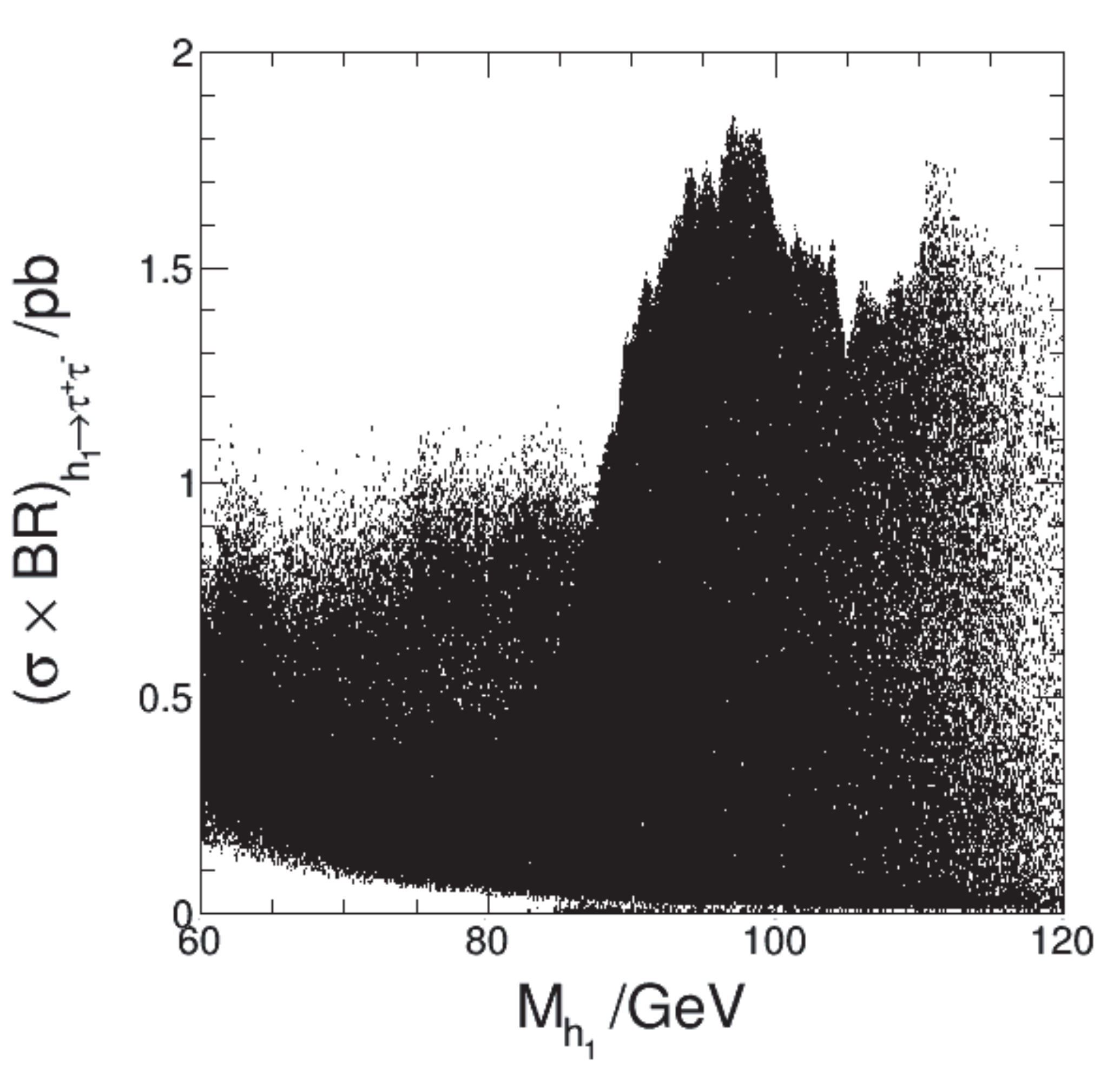}\quad
\includegraphics[width=0.3\textwidth,height=0.3\textwidth]{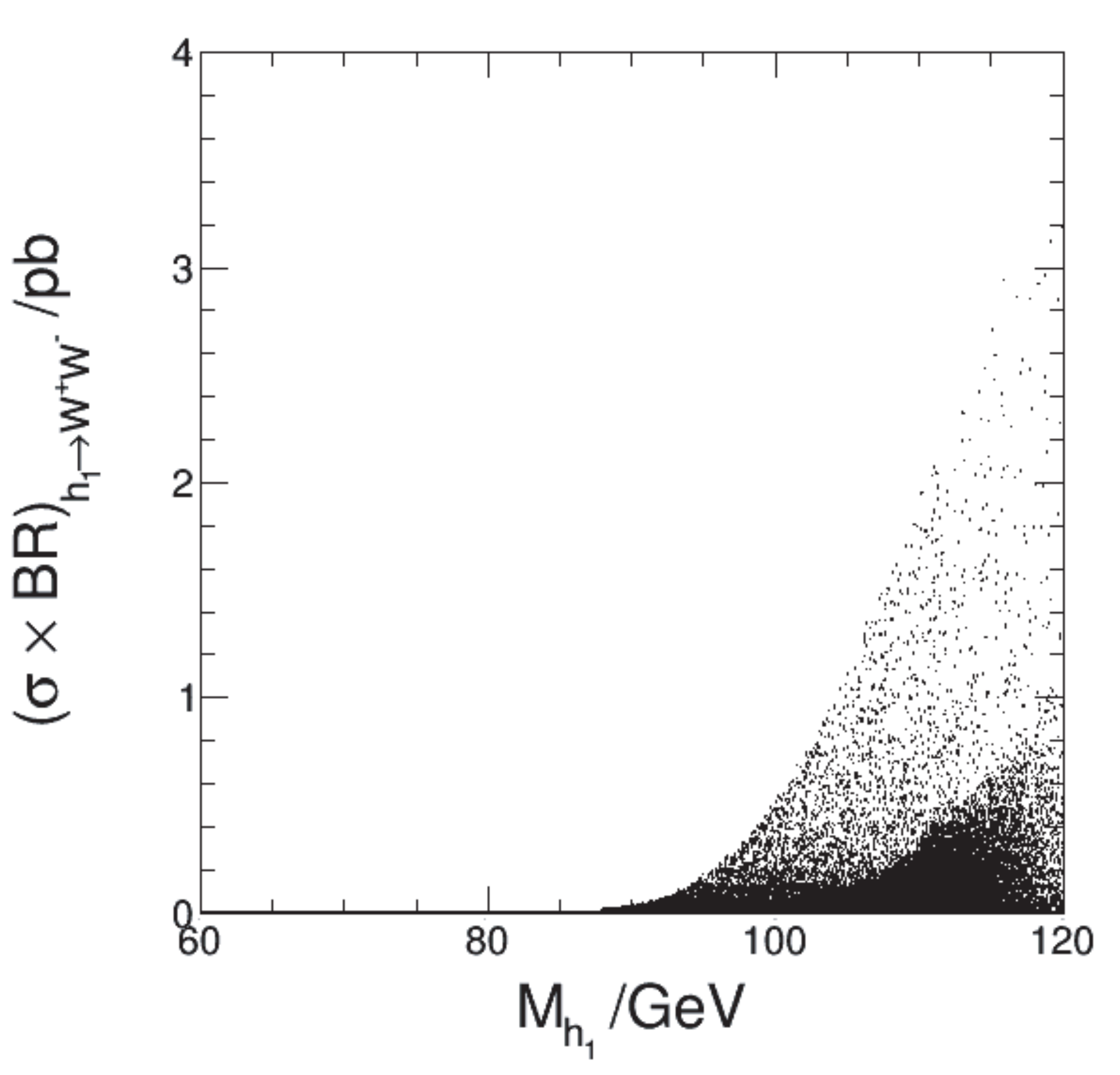} \\
\includegraphics[width=0.3\textwidth,height=0.3\textwidth]{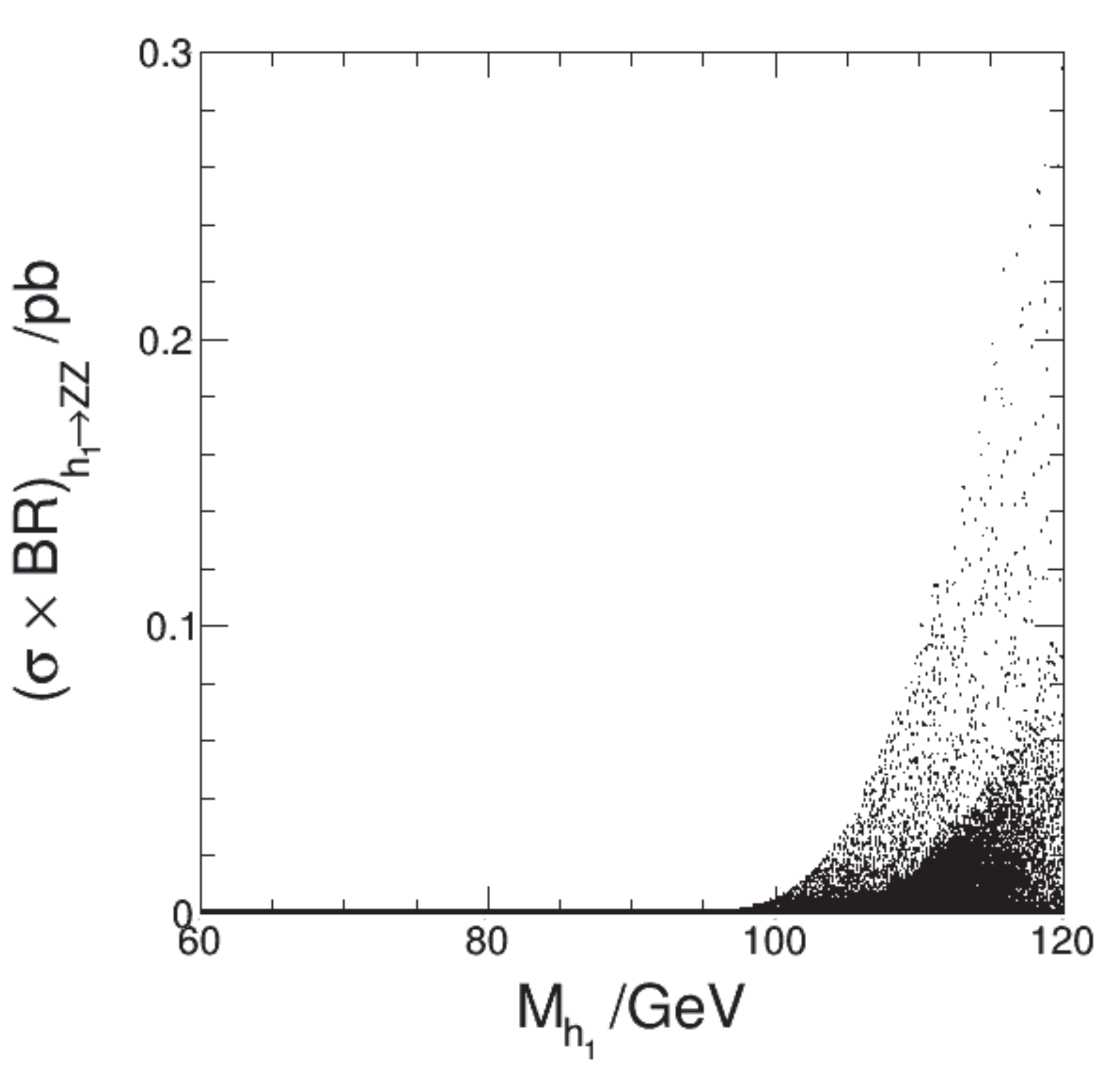}\quad
\includegraphics[width=0.3\textwidth,height=0.3\textwidth]{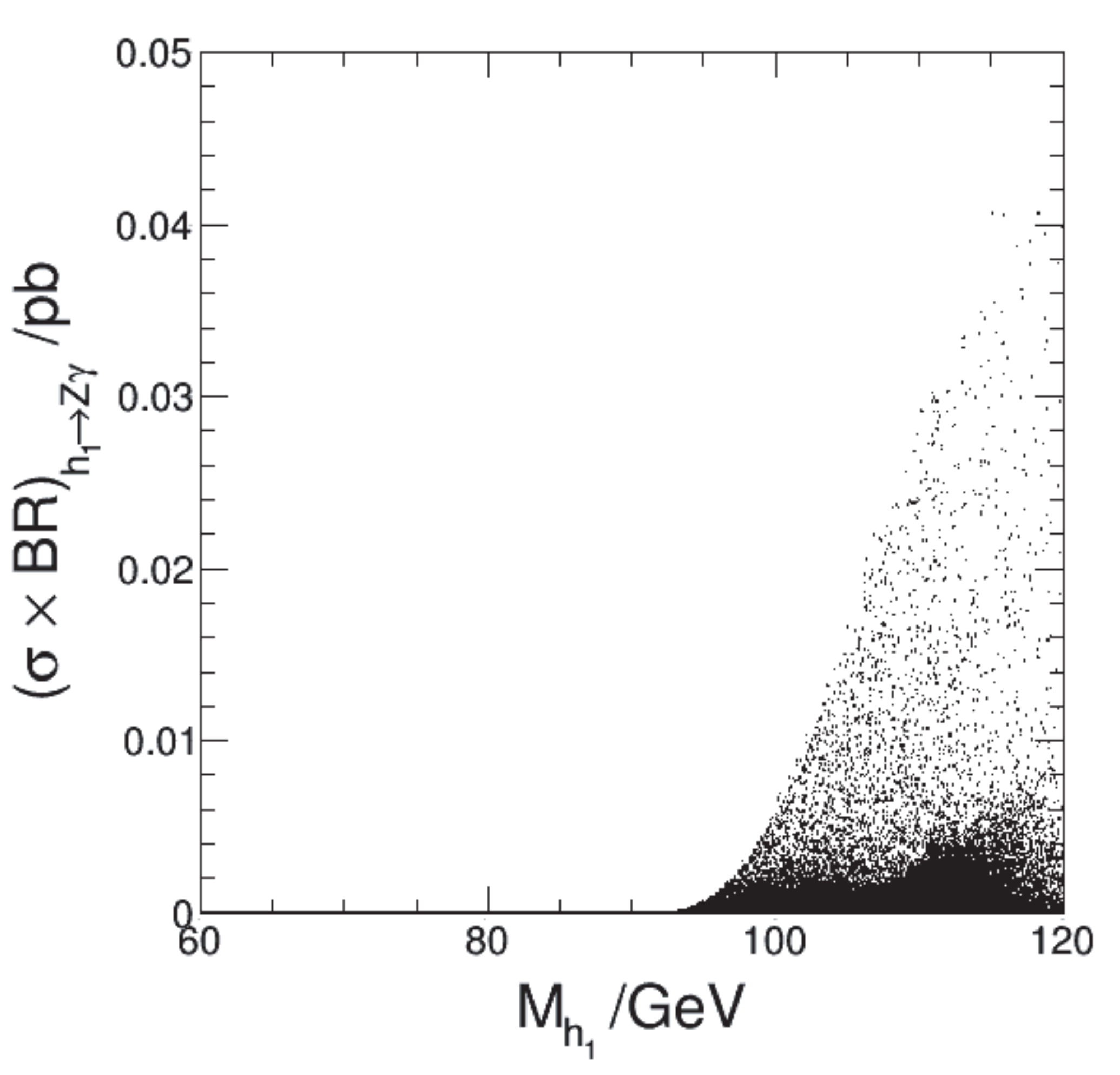}\quad
\includegraphics[width=0.3\textwidth,height=0.3\textwidth]{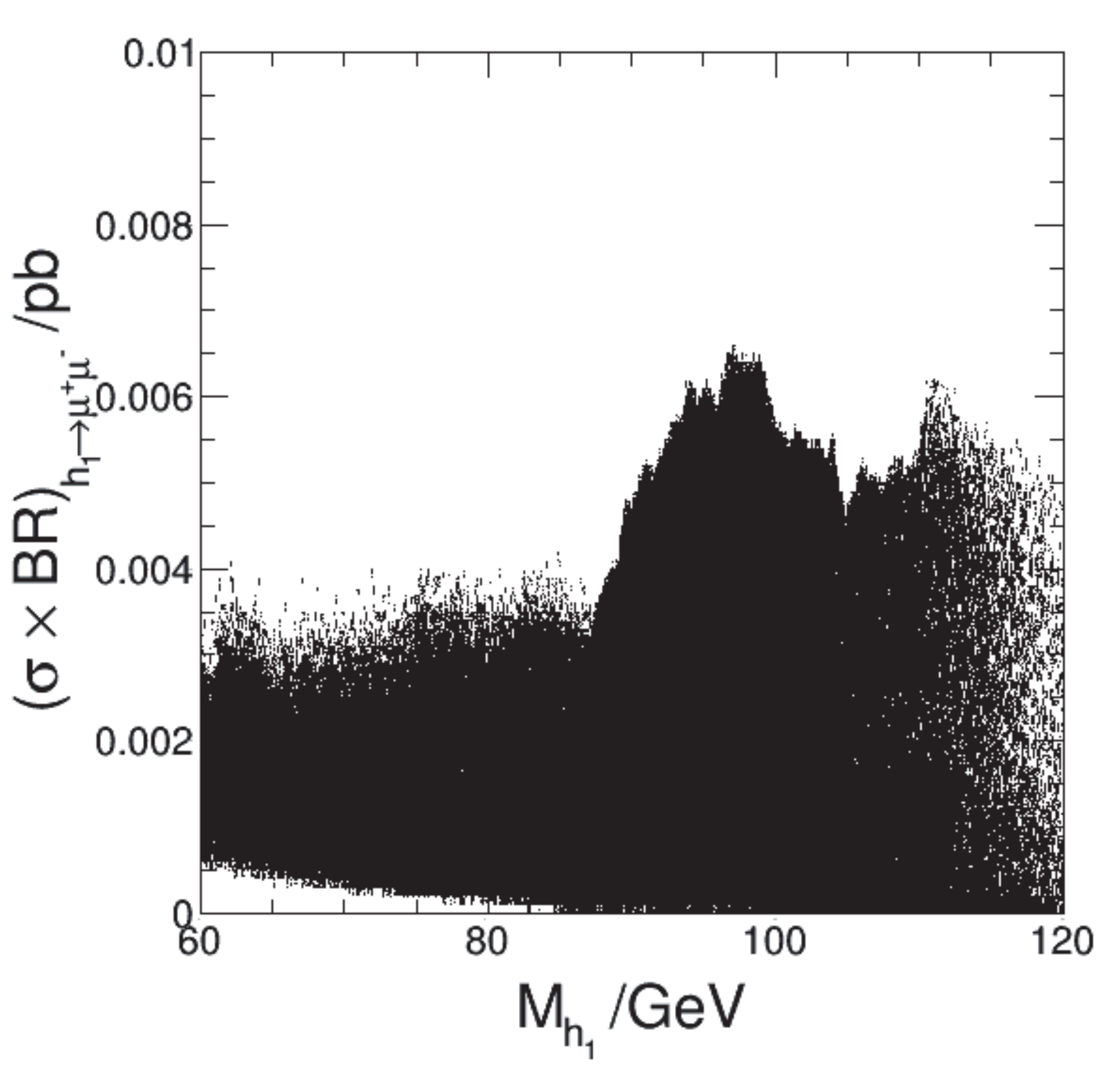}
\figcaption{\label{fig:h1mu_mass} Signal rates as functions of the $h_{1}$ mass for other interesting decay channels: $(\sigma \times BR)_{h_{1}\rightarrow b\bar{b} }$ (top left), $(\sigma \times BR)_{h_{1}\rightarrow\tau^{+}\tau^{-}}^{ggh+tth}$ (top middle), $(\sigma \times BR)_{h_{1}\rightarrow W^{+}W^{-}}^{vbf+vh}$ (top right), $(\sigma \times BR)_{h_{1}\rightarrow ZZ}^{ggh+tth}$ (bottom left), $(\sigma \times BR)_{h_{1}\rightarrow Z\gamma}^{vbf+vh}$ (bottom middle), and $(\sigma \times BR)_{h_{1}\rightarrow\mu^{+}\mu^{-}}^{vbf+vh}$ (bottom right). }
\end{figure*}

In addition, we also checked the production rates of other interesting decay channels, including $b\bar{b}$, $\tau^{+}\tau^{-}$, $W^{+}W^{-}$,
$ZZ$, $Z\gamma$, and $\mu^{+}\mu^{-}$, to investigate the discovery potentials of $h_{1}$ in these channels.
Fig.~\ref{fig:h1mu_mass} shows the production rates in picobarns (pb) for $h_{1}$ decaying into $b\bar{b}$, $\tau^{+}\tau^{-}$, $W^{+}W^{-}$,
$ZZ$, $Z\gamma$, and $\mu^{+}\mu^{-}$, as functions of its mass $M_{h_{1}}$. The $h_{1}\rightarrow b\bar{b}$, $h_{1}\rightarrow\tau^{+}\tau^{-}$,
and $h_{1}\rightarrow\mu^{+}\mu^{-}$ decay have tight correlations on the branching ratios, although with different values.
Therefore, the shapes of their production rates as functions of $M_{h_{1}}$ are similar.
In addition, the similar shapes of production rates of $h_{1}\rightarrow W^{+}W^{-}$,
$h_{1}\rightarrow ZZ$ and $h_{1}\rightarrow Z\gamma$ are a result of the tight correlations on the branching ratios.
Among these decay channels, the signal rate of $h_{1}\rightarrow b\bar{b}$ is reasonably large, as the rates can reach
up to 18~pb with $M_{h_{1}}$ at around 95~GeV. For the $h_{1}\rightarrow W^{+}W^{-}$,
$h_{1}\rightarrow ZZ$, and $h_{1}\rightarrow Z\gamma$ channels, the signal rates decrease with decreasing $M_{h_{1}}$.
For the $b\bar{b}$ and $\tau^{+}\tau^{-}$ final states in the investigated mass range, the signal rates are sufficiently large
that it is very possible to detect $h_{1}$ by experiments at the LHC via these two channels.

\vspace{-0.2mm}
\section{Results for a lighter pseudo-scalar Higgs boson} \label{sec:pseudo}
\vspace{-0.2mm}

As the kinematic behavior of the two photons coming from the decay of a pseudo-scalar particle is very similar to that resulting
from a scalar particle~\cite{Artoisenet:2013puc}, we can directly apply the CMS study as in the scalar case to constrain a possible light pseudo-scalar.
Because the mass of the heavier pseudo-scalar $a_{2}$ from NMSSM scans is considerably larger than that of the Higgs observed at the LHC, we focus on the lightest pseudo-scalar $a_{1}$ in the following. From the random scans after the phenomenological constraints and the mass and signal strength constraints on $h_{2}$ have been imposed, the mass distributions of
the lightest pseudo-scalar $a_{1}$ versus the scalar Higgs bosons $h_{2}$ are shown in the top left panel of Fig.~\ref{fig:a1}.
Then, about 187,000 points are selected after the constraint of $a_{1}$ within the mass range from 60~GeV to 120~GeV has been imposed.

The lightest CP-odd Higgs boson $a_{1}$ primarily decays to fermions, owing to the absence of tree-level couplings with gauge bosons. As
shown in the top-right panel of Fig.~\ref{fig:a1}, it decays dominantly to $b\bar{b}$ with BR $\sim$90\% for the low mass range.
For $a_{1}\rightarrow\gamma\gamma$ with $a_{1}$ in the mass range of 60~GeV to 120~GeV, the BR is less than $7\times10^{-4}$ for all the selected points.
As shown in the bottom left panel, the signal rates of $a_{1}\rightarrow\gamma\gamma$ for all the selected points in the combined ggh and tth production mode are lower than 0.3 $fb$, which is far below the CMS-observed upper limits, as the red line shows in the middle panel of Fig.~\ref{fig:h1ggmu_mass}.
In addition, for the combined vbf and vh production mode the signal rates of $a_{1}\rightarrow\gamma\gamma$
with the points shown in the bottom right panel of Fig.~\ref{fig:a1} are also far below the CMS-observed upper limit on the production
cross-section times the branching ratio, as the red line shows in the right panel of Fig.~\ref{fig:h1ggmu_mass}. Therefore, we conclude that CMS had no
sensitivity to a light pseudo-scalar in the diphoton final state with the data collected in the year 2016.

We have also checked the production rates for other interesting decay channels of $a_{1}$, to investigate the discovery potentials of $a_{1}$ in these channels.
Fig.~\ref{fig:a1mu_mass} shows the production rates in femtobarns (fb) for $a_{1}$ decaying into $b\bar{b}$, $\tau^{+}\tau^{-}$, $W^{+}W^{-}$,
$ZZ$, $Z\gamma$, and $\mu^{+}\mu^{-}$ as functions of its mass $M_{a_{1}}$.
As expected, $a_{1}\rightarrow b\bar{b}$ is the dominant decay channel, with signal rates of up to about 3800 fb, and
$h_{1}\rightarrow\tau^{+}\tau^{-}$ is the sub-dominant decay channel, with signal rates of up to about 300 fb.
For the $b\bar{b}$ decay of $a_{1}$, the cross section is sufficiently large to search for $a_{1}$ at the LHC if the large backgrounds can be well suppressed.
As for the top quark pair final states, it is also possible to detect a low-mass $a_{1}$ in this channel.
Considering the BRs of the cascade decays of $W$ and $Z$,
it will be difficult to search for $a_{1}$ with the $W^{+}W^{-}$ and $ZZ$ channels with all the LHC Run2 data.
It is also difficult to search for $a_{1}$ with $h_{1}\rightarrow\mu^{+}\mu^{-}$, owing to the small signal rates
and the acceptance times the selection efficiency for the signal events, which is $\sim$50\% for a 125~GeV SM Higgs boson~\cite{CMS:2017qgo,Aaboud:2017ojs}.
For the lower signal rates of $Z\gamma$ decay, it is impossible to search for $a_{1}$ in the $Z\gamma$ channel for all LHC Run2 data.

\end{multicols}

\begin{center}
\centering
\includegraphics[width=0.3\textwidth,height=0.3\textwidth]{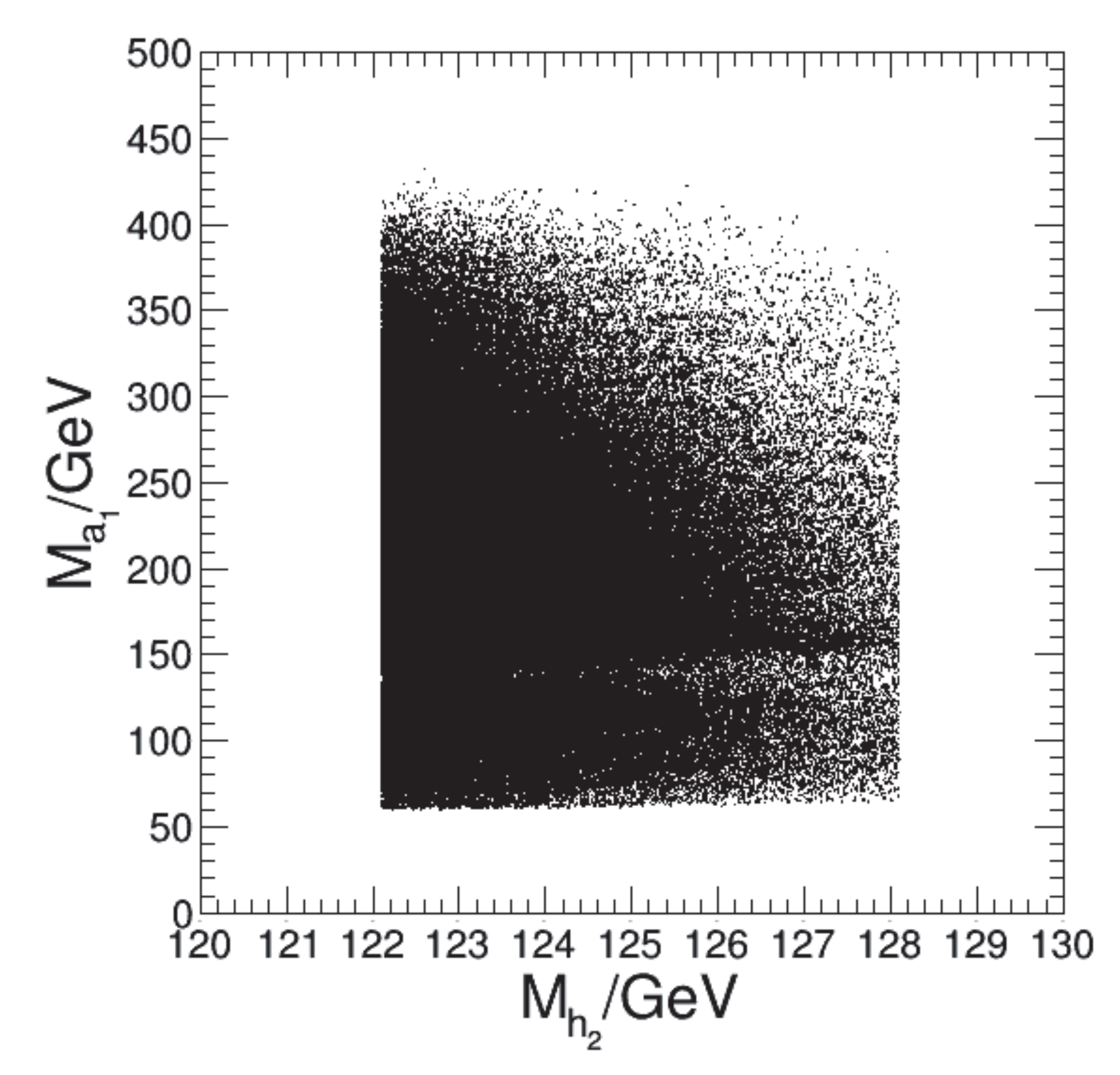}\quad
\includegraphics[width=0.3\textwidth,height=0.3\textwidth]{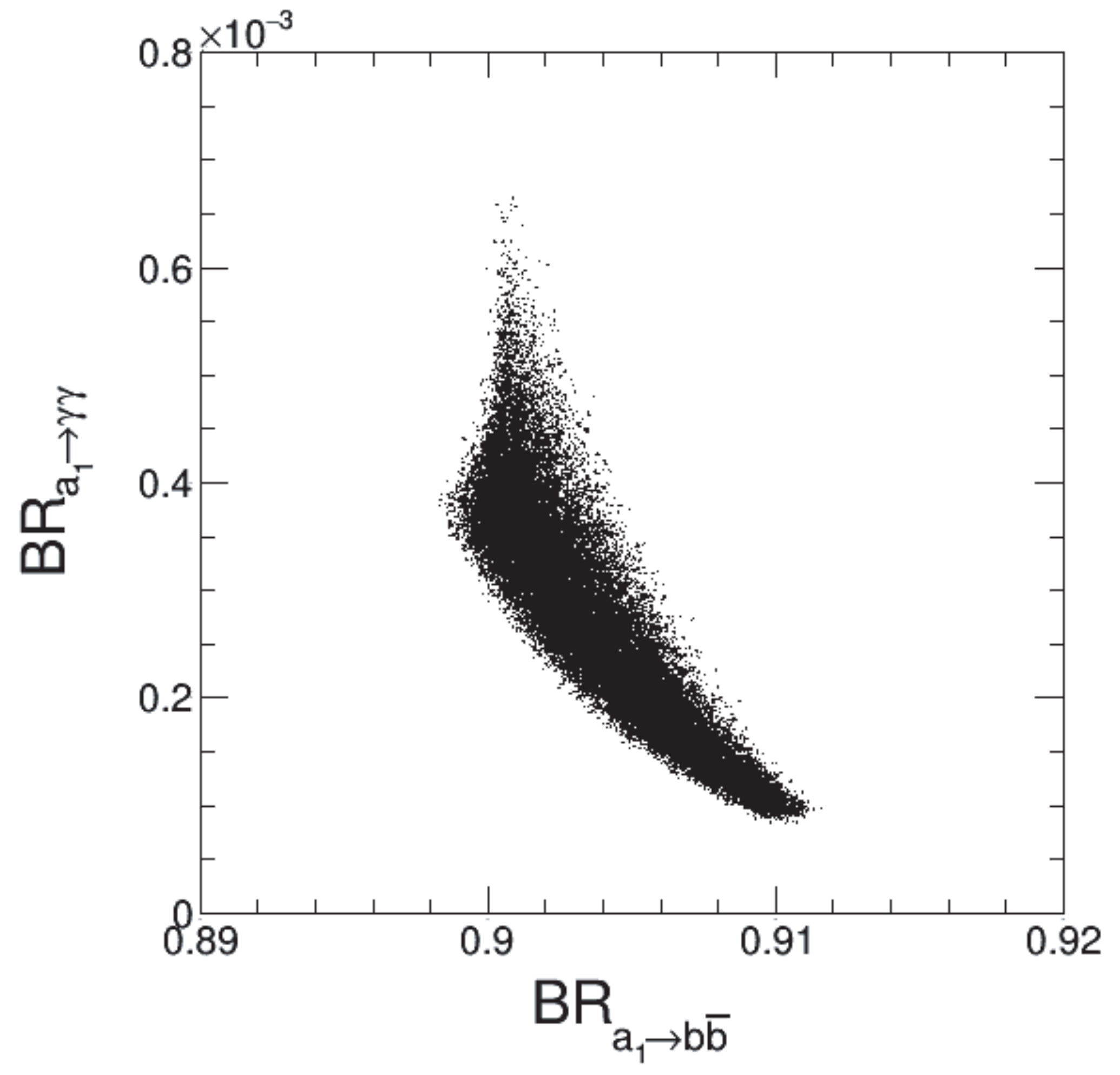} \\
\includegraphics[width=0.3\textwidth,height=0.3\textwidth]{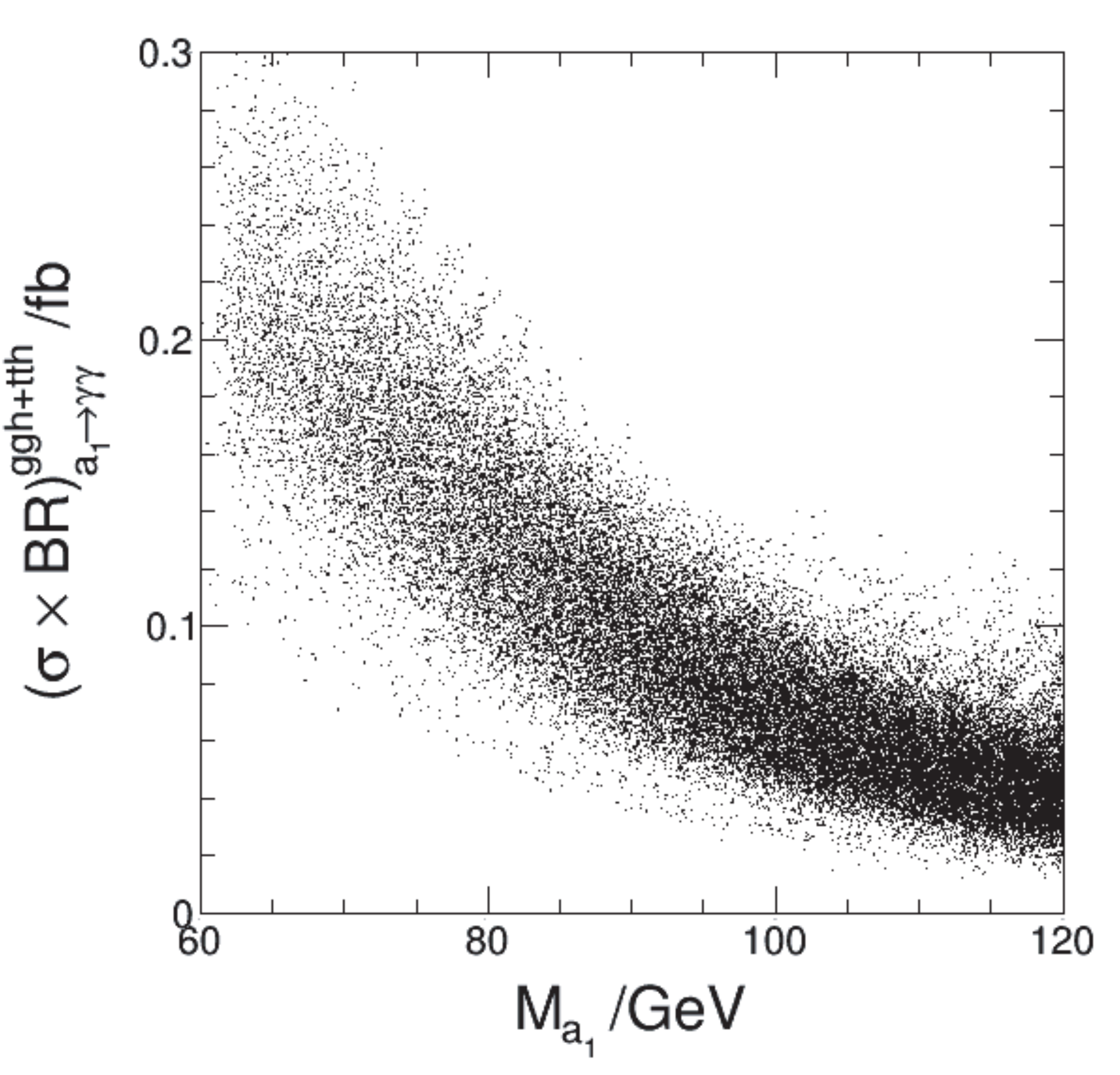}\quad
\includegraphics[width=0.3\textwidth,height=0.3\textwidth]{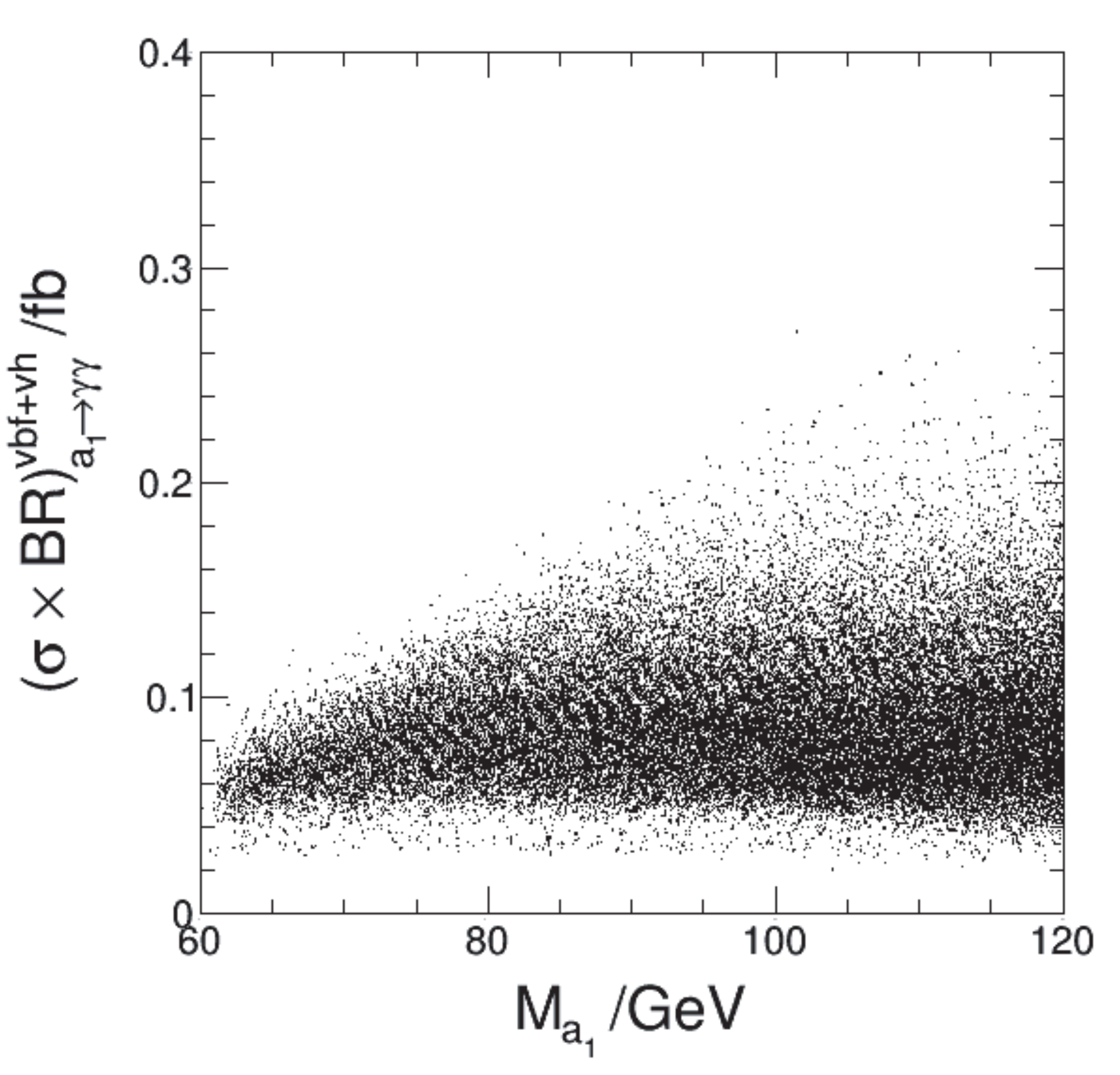}
\figcaption{\label{fig:a1} Mass spectrum of the lightest pseudo-scalar Higgs bosons $a_{1}$ versus $h_{2}$ (top left panel) after the phenomenological constraints and constraints on $h_{2}$ have been imposed,
distributions of the branching ratios of $a_{1}\rightarrow b\bar{b}$ versus $a_{1}\rightarrow\gamma\gamma$ (top right) and the signal rates of $a_{1}\rightarrow\gamma\gamma$ versus the mass of $a_{1}$ for different combined production modes with $(\sigma \times BR)_{a_{1}\rightarrow\gamma\gamma}^{ggh+tth}$ for ggh+tth (bottom left), and $(\sigma \times BR)_{a_{1}\rightarrow\gamma\gamma}^{vbf+vh}$ for vbf+vh (bottom right) after the further mass constraint on $a_{1}$ has been imposed. }
\end{center}

\begin{multicols}{2}

\section{Conclusions} \label{sec:conclusions}

Following the discovery of the Higgs boson with a mass of approximately
125~GeV at the LHC, many studies from both the theoretical and experimental
viewpoints have been performed to search for a new Higgs Boson that is lighter than 125~GeV.
The search for such a lighter Higgs Boson
represents one of the most important avenues for probing the possible structure of physics beyond the Standard
Model. In this paper, we explored the possibility of constraining a lighter neutral scalar Higgs boson $h_{1}$ and a lighter pseudo-scalar Higgs boson $a_{1}$ in the
Next-to-Minimal Supersymmetric Standard Model by restricting the
next-to-lightest scalar Higgs boson $h_{2}$ to be the LHC observed
Higgs boson after the phenomenological constraints and constraints from experimental measurements have been imposed.
Such a lighter particle is not yet completely excluded by the latest results of the search for a lighter Higgs boson
with the diphoton decay channel from the LHC data collected by the CMS detector at 13~TeV.
For a lighter neutral scalar $h_{1}$, we can expect to exclude some NMSSM region in the parameter space.
While the latest CMS results shows no sensitivity to a light pseudo-scalar in the diphoton final state,
our results show that some new constraints on the Next-to-Minimal Supersymmetric Standard Model could be obtained at the LHC if such a search
is performed by the experimental collaborations with additional data in the future.
The discovery potentials for other interesting decay channels of such a lighter neutral scalar or pseudo-scalar particle have also been discussed.
For the $b\bar{b}$ and $\tau^{+}\tau^{-}$ final states of both $h_{1}$ and $a_{1}$  in the investigated mass range,
it is possible to detect such lighter particles by the experiments at the LHC.\\

\textit{The authors would like to thank Ulrich Ellwanger and Cyril Hugonie for helpful
discussions.}

\end{multicols}

\begin{center}
\centering
\includegraphics[width=0.3\textwidth,height=0.3\textwidth]{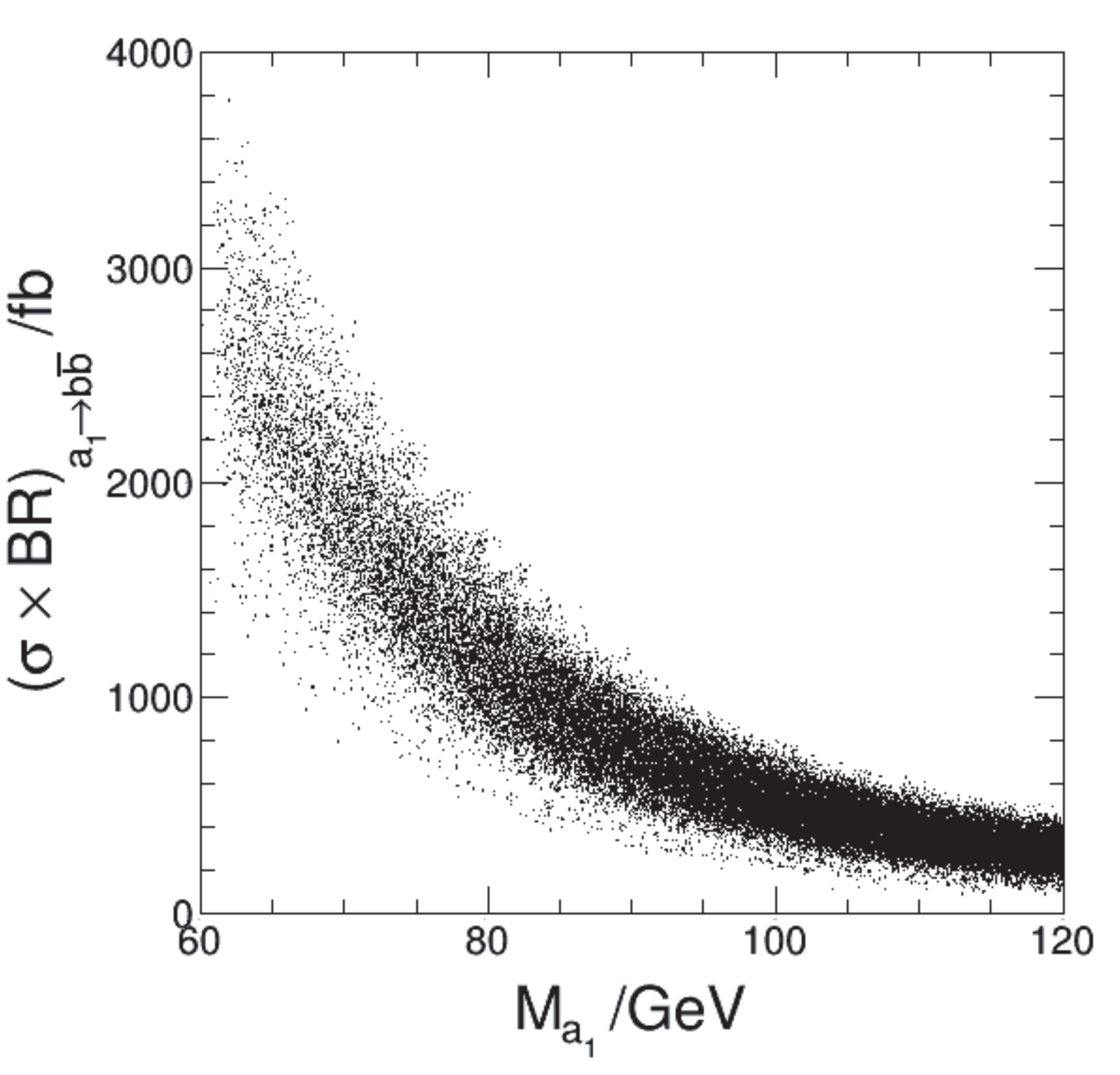}\quad
\includegraphics[width=0.3\textwidth,height=0.3\textwidth]{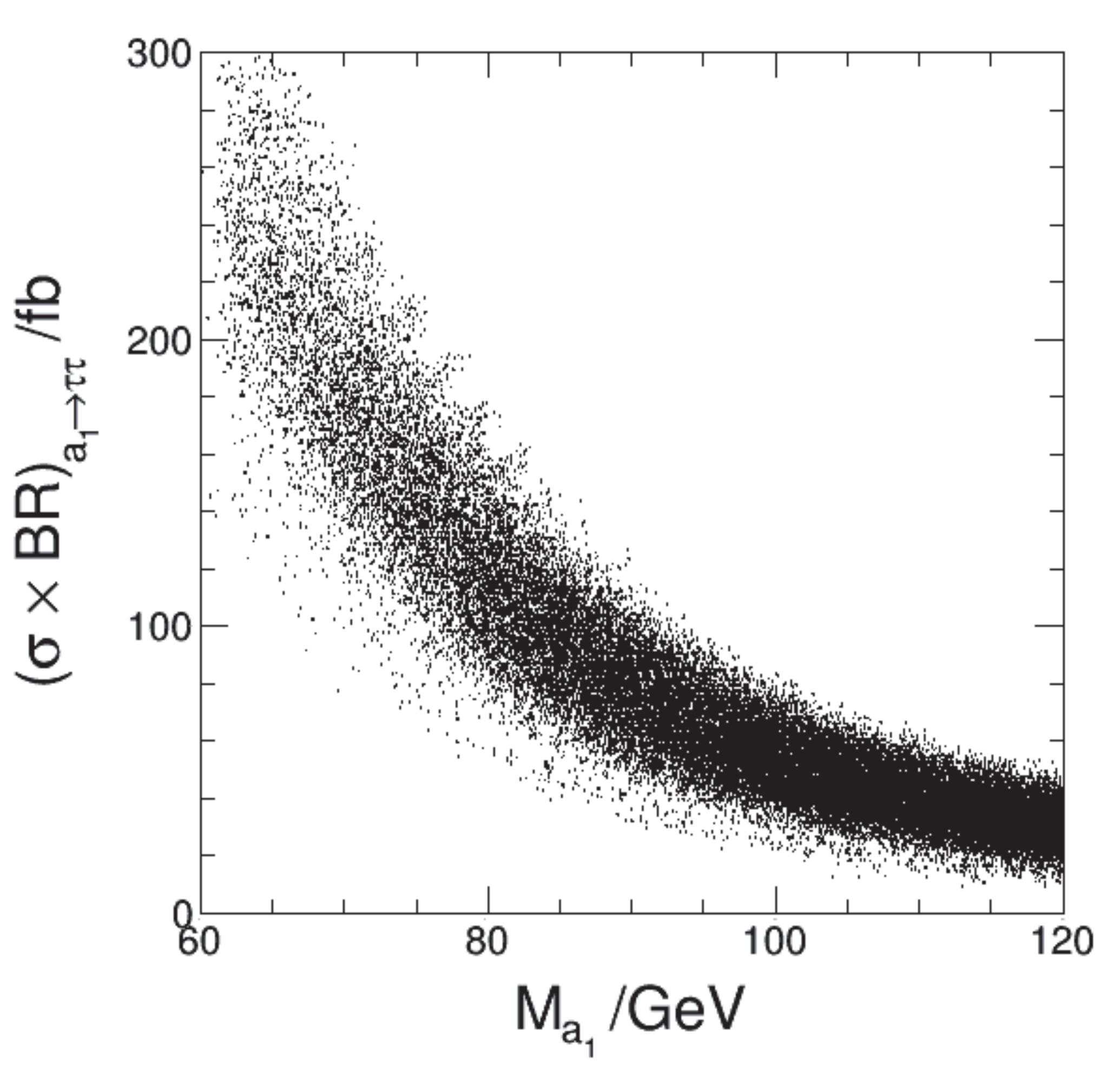}\quad
\includegraphics[width=0.3\textwidth,height=0.3\textwidth]{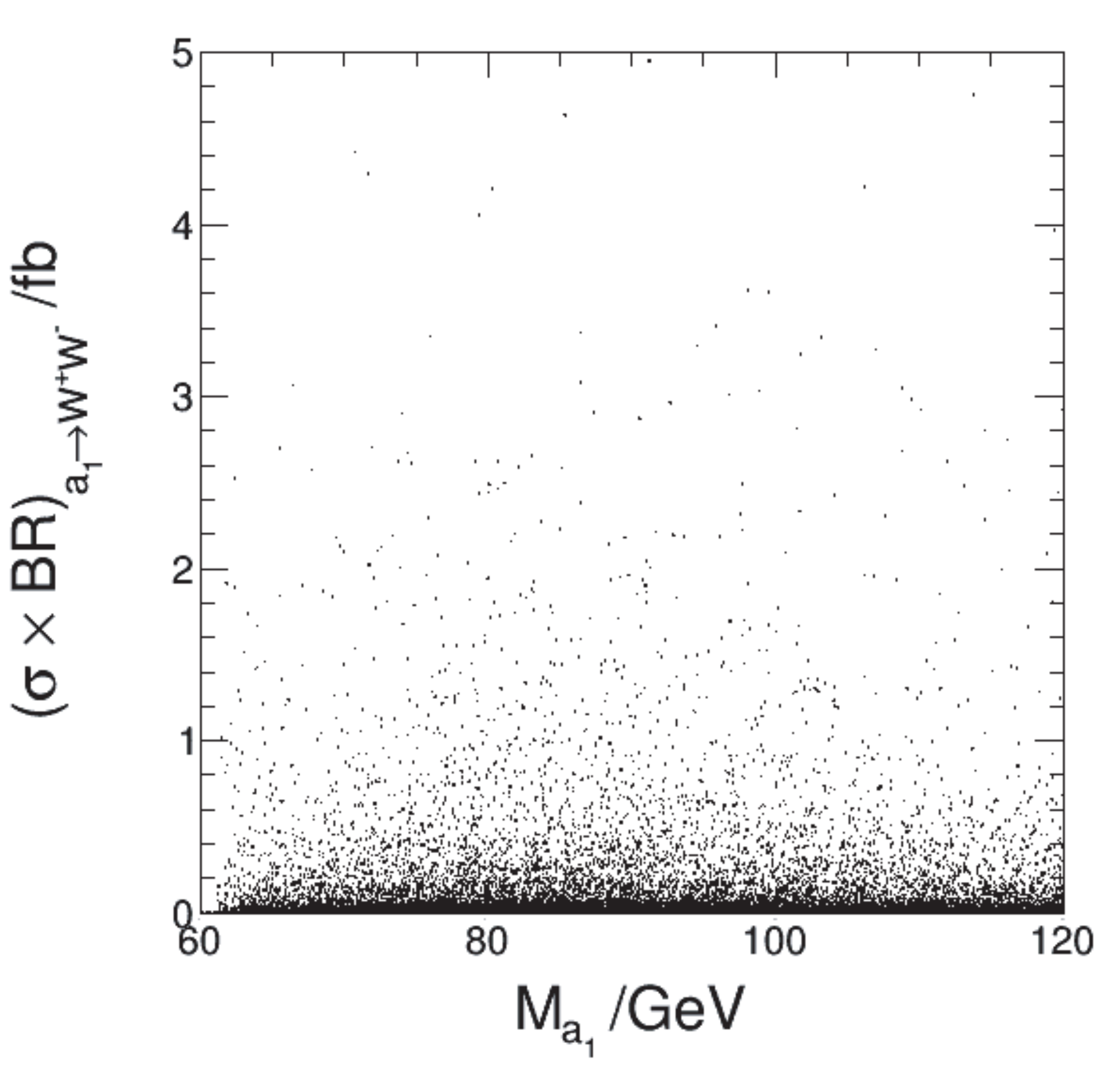} \\
\includegraphics[width=0.3\textwidth,height=0.3\textwidth]{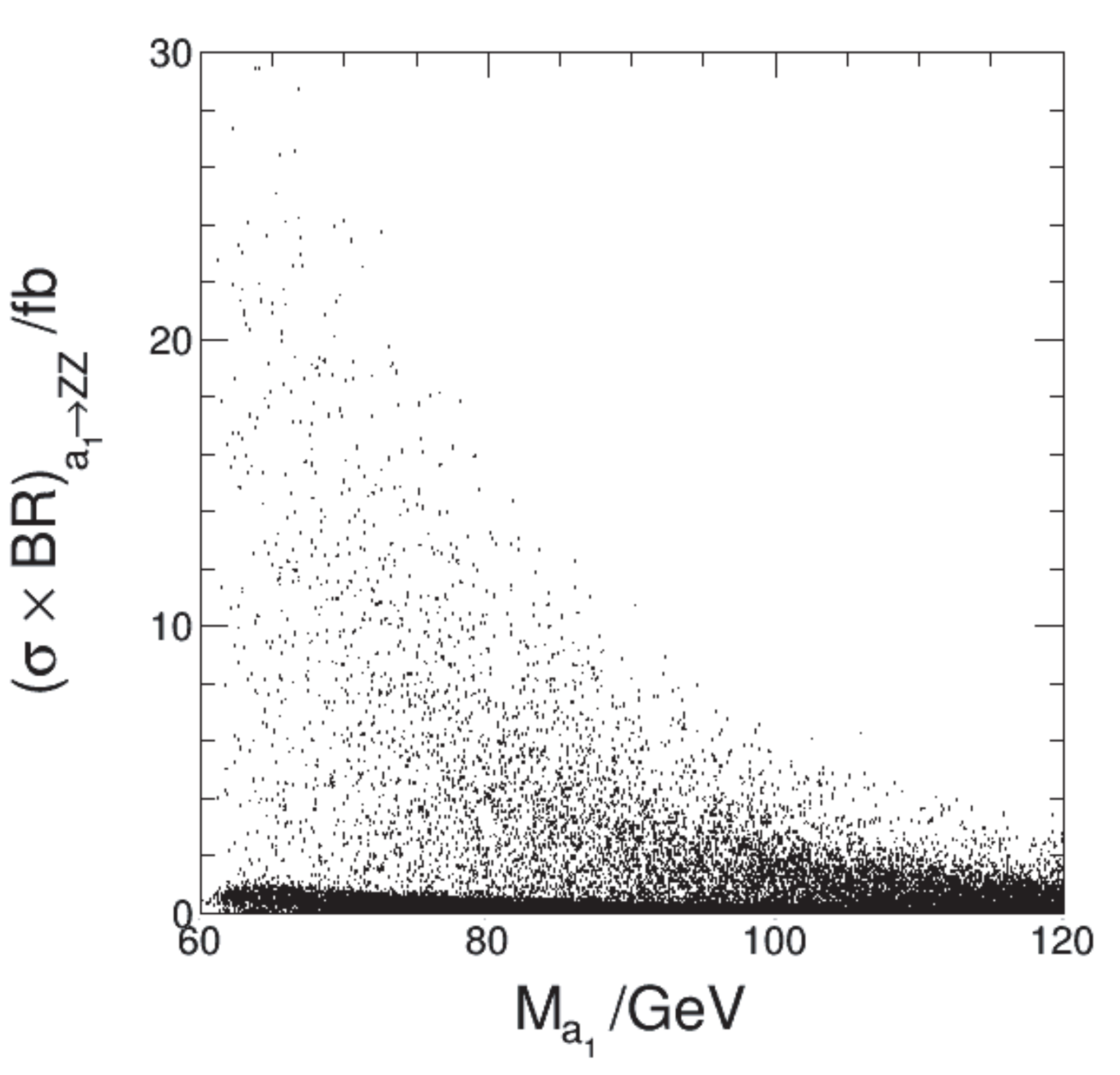}\quad
\includegraphics[width=0.3\textwidth,height=0.3\textwidth]{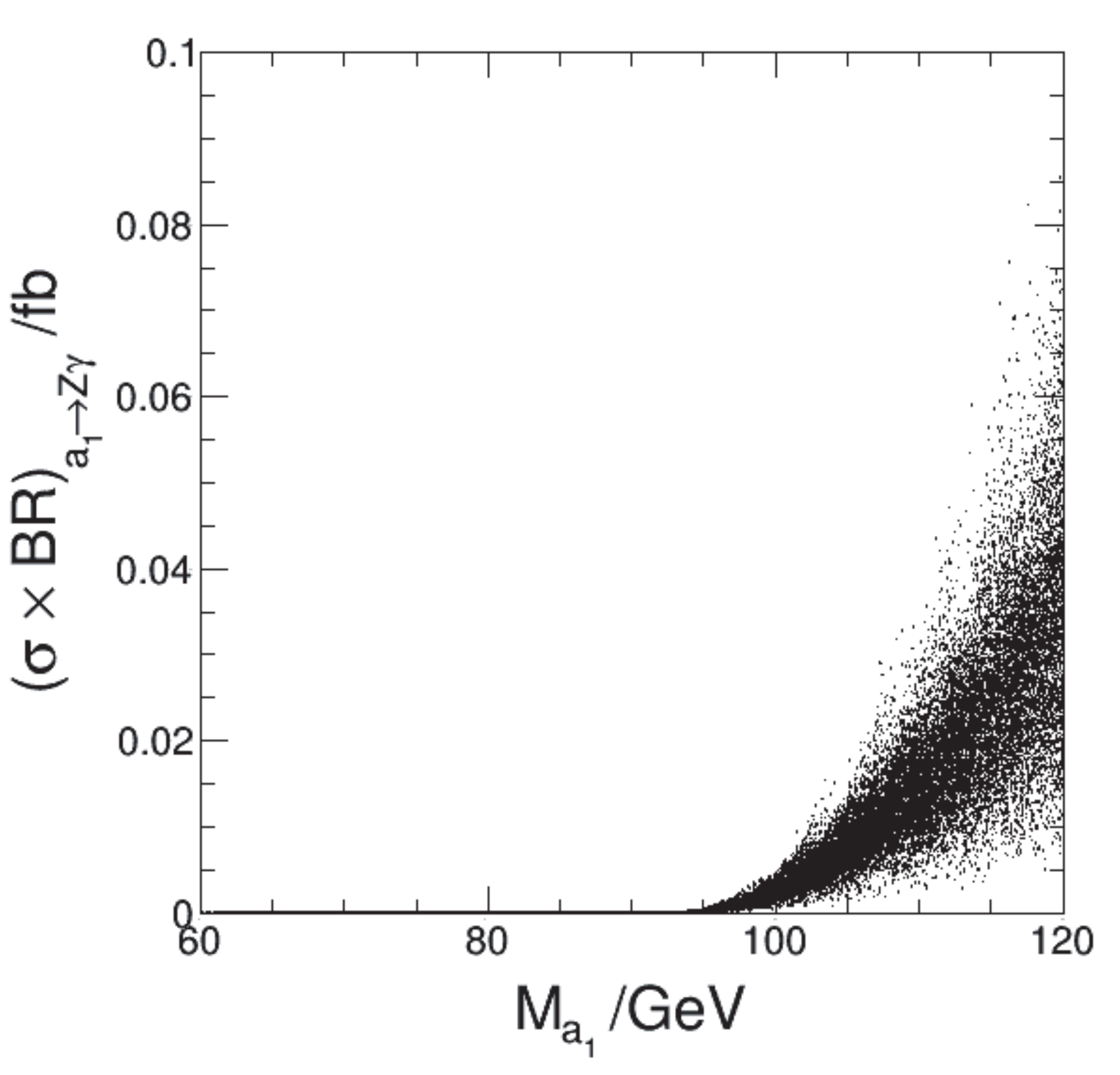}\quad
\includegraphics[width=0.3\textwidth,height=0.3\textwidth]{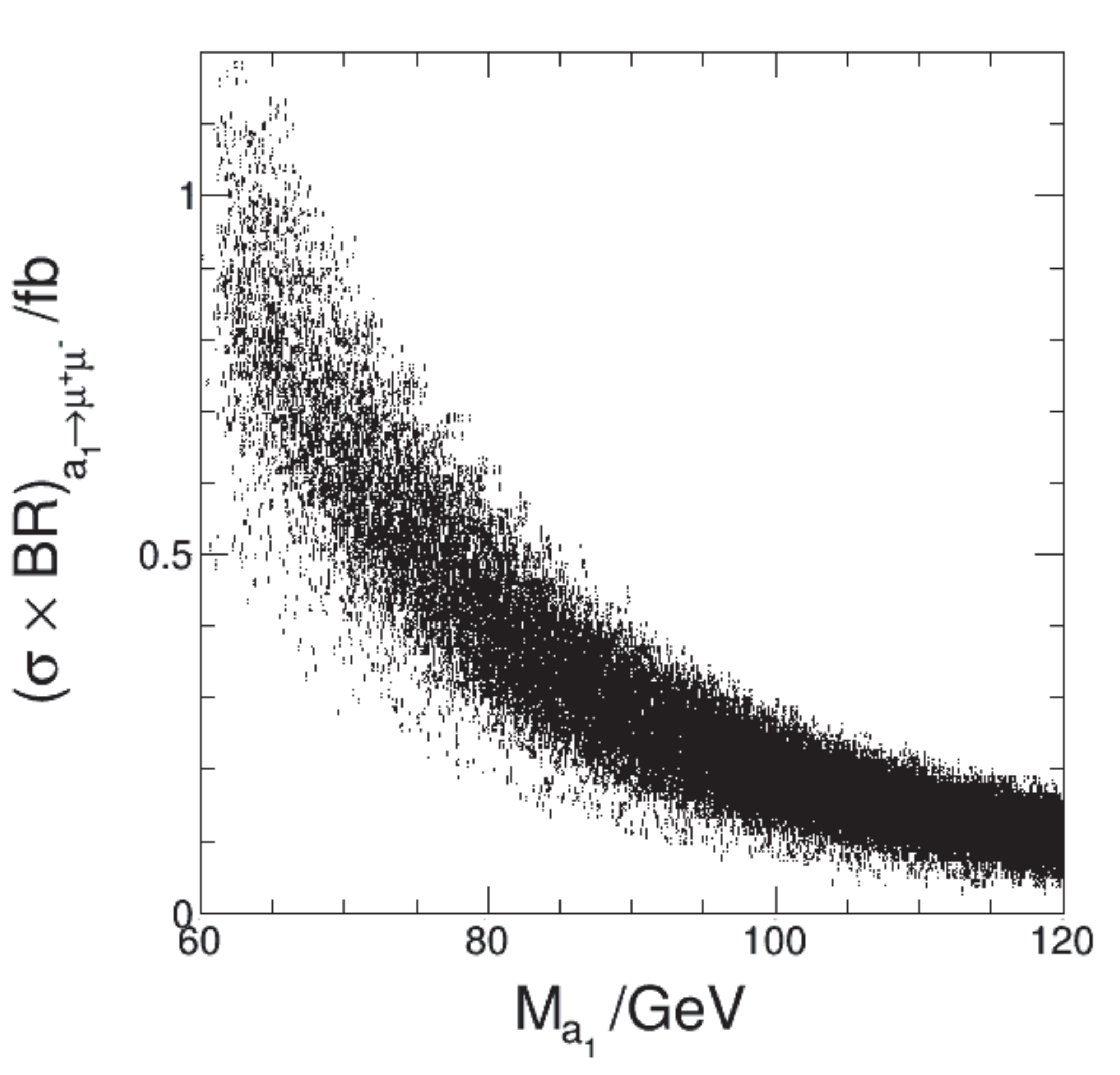}
\figcaption{\label{fig:a1mu_mass} Signal rates as functions of the $a_{1}$ mass for other interesting decay channels: $(\sigma \times BR)_{a_{1}\rightarrow b\bar{b} }$ (top left), $(\sigma \times BR)_{a_{1}\rightarrow\tau^{+}\tau^{-}}$ (top middle), $(\sigma \times BR)_{a_{1}\rightarrow W^{+}W^{-}}$ (top right), $(\sigma \times BR)_{a_{1}\rightarrow ZZ}$ (bottom left), $(\sigma \times BR)_{a_{1}\rightarrow Z\gamma}$ (bottom right), and $(\sigma \times BR)_{a_{1}\rightarrow\mu^{+}\mu^{-}}$ (bottom right). }
\end{center}

\vspace{-2.5mm} \centerline{\rule{80mm}{0.1pt}} \vspace{1mm}

\begin{multicols}{2}

\end{multicols}

\clearpage

\end{CJK*}
\end{document}